\documentstyle[12pt,prc,aps,epsfig,graphicx]{revtex}

\def\mpi{m_{\pi}}

\def\beq{\begin{equation}}
\def\eeq{\end{equation}}
\def\bea{\begin{eqnarray}}
\def\eea{\end{eqnarray}}

\def\sla#1{#1 \hspace{-2.7mm} \slash}
\def\slas#1{#1 \hspace{-1.9mm} \slash}
\def\bold#1{\mbox{\bf #1}}
\def\boldp#1{\bold{#1{$^{\prime}$}}}
\def\boldh#1{\bbox{\hat{\mbox{\bf #1}}}}
\def\boldhp#1{\bbox{\hat{\mbox{\bf #1}}^{\prime}}}
\def\boldhpp#1{\bbox{\hat{\mbox{\bf #1}}^{\prime \prime}}}

\def\half{\mbox{\small{$\frac{1}{2}$}}}
\def\thalf{\mbox{\small{$\frac{3}{2}$}}}

\def\lag{{\cal L}}

\def\CF#1#2#3#4{#1 {\bf #2}, #4 (#3).}  
\def\CFA#1#2#3#4{#1 {\bf #2}, #4 (#3);}

\def\ib {{\it ibid.}}

\def\jpg {J.~Phys.~G}

\def\ncim {Nuovo~Cim.}
\def\np {Nucl.~Phys.}
\def\prr {Phys.~Rev.}
\def\prc {Phys.~Rev.~C}
\def\prd {Phys.~Rev.~D}

\def\plb {Phys.~Lett.~B}
\def\prl {Phys.~Rev.~Lett.}
\def\prep {Phys.~Rep.}

\newcommand{\solid}{\protect\rule[1mm]{6mm}{.1mm}}
\newcommand{\dash}{\protect\rule[1mm]{2mm}{.1mm}\hspace*{-0.05cm}
\protect\rule[1mm]{2mm}{.1mm}}

\newcommand{\dashdash}{\protect\rule[1mm]{3mm}{.1mm}\hspace*{-0.05cm}
\protect\rule[1mm]{0.8mm}{.1mm}}

\newcommand{\dashes}{\protect\rule[1mm]{1mm}{.1mm}\protect\hspace*{-0.04cm}
\protect\rule[1mm]{1mm}{.1mm}}

\newcommand{\dashdot}{\protect\rule[1mm]{2mm}{.1mm}\hspace{1mm}$\cdot$}

\newcommand{\dotss}{\protect$\cdot\cdot\cdot$}

\tightenlines

\begin{document}

\title{Solution of the Bethe-Salpeter equation for \\ pion-nucleon scattering}

\author{A. D. Lahiff and I. R. Afnan}

\address{Department of Physics,
        The Flinders University of South Australia, \\
        GPO Box 2100, Adelaide 5001, Australia}

\date{\today}

\maketitle

\begin{abstract}
A relativistic description of pion-nucleon scattering based on the
four-dimensional Bethe-Salpeter equation is presented. The kernel of
the equation consists of $s$- and $u$-channel nucleon and
$\Delta (1232)$ pole diagrams, as well as $\rho$  and $\sigma$ exchange
in the $t$-channel. The Bethe-Salpeter equation is solved by means 
of a Wick rotation, and good fits are obtained to the $s$- and $p$-wave 
$\pi N$ phase shifts up to 360 MeV pion laboratory energy. The coupling
constants determined by the fits are consistent with the 
commonly accepted values in the literature.

\end{abstract}

\pacs{13.75.Gx,11.10.St,24.10.Jv,21.45.+v}

\section{Introduction}\label{sec.1}

Pion-nucleon ($\pi N$) scattering is an important example of a
strong interaction, and as such plays a significant role
in many nuclear reactions involving pions, the most interesting
example in recent years being pion photoproduction. 
It is generally accepted that the fundamental theory of strong 
interactions is Quantum Chromodynamics (QCD), and therefore a theory 
describing $\pi N$ scattering should ideally  be derived from QCD. 
However, due to the non-perturbative nature of confinement, QCD has
not been amiable to solutions for low and intermediate energies, and
hence it is necessary to use an effective theory.
In principle the effective theory
should be as close as possible to the fundamental one, and so should satisfy
the same symmetries, in particular chiral symmetry, which is known
to be important for low-energy physics.
Therefore, in place of the QCD Lagrangian
a chirally-invariant hadronic Lagrangian is used, where the 
degrees of freedom are mesons and baryons rather than quarks and gluons. 
For low energies it is expected that the detailed quark structure of hadrons
is relatively unimportant, and that it is only at very high
energies that explicit quark degrees of freedom are essential.
The great success of meson-exchange models for nucleon-nucleon ($NN$)
scattering in the description of $NN$ phase shifts is an example of this.

A number of dynamical models of $\pi N$ scattering have been 
developed over the past few years. Most begin with a potential 
which is iterated in a 
Lippman-Schwinger-type equation to give the scattering amplitude, from
which the phase shifts and observables are obtained.
This method ensures that two-body unitarity is respected, and that 
multiple scattering effects are taken into account.
The simplest models use 
separable potentials~\cite{sep1,sep2,sep3,pncre1,pncre2}, in which 
the parameters have no physical meaning, and
furthermore, different sets of parameters are used in each
partial wave. While these
models provide  good descriptions of the $\pi N$ phase shifts,
they provide no information about the interaction process.
An alternative 
is to derive a potential from a Lagrangian
which describes the couplings between the various mesons and baryons.
In tree-level models~\cite{ols75,bof91,goud94,mosel,kor98},
the potential is unitarized using the
$K$-matrix approximation, which relies on the assumption that for
low energies the $K$-matrix is equal to the potential.
While this method can provide a good description of the $\pi N$ data at low energies,
in order to cover a larger region of energies and to be able to
investigate the nature of resonances, the potential must be iterated to all orders.
This has been done 
recently~\cite{pj91,leeyang91,gross93,leeyang94,schutz94,schutz98,tjon98,tjon98a}
in models based on three-dimensional (3-D) approximations to the Bethe-Salpeter 
equation~\cite{bse}. Phenomenological form factors are required at the vertices in order to
provide convergence, and so in general more free parameters are needed
than in tree-level models, which can give good results without any
form factors~\cite{goud94}.

There are several other approaches to $\pi N$ scattering in the
literature which give results of a similar quality.
The meson-exchange model of J{\"a}de~\cite{jade} uses the solitary boson
exchange method to regularize self-energy diagrams instead of using
form factors, and gives good descriptions of both $\pi N$ and $NN$ 
scattering with the same set of parameters. Sato and Lee constructed a 
meson-exchange model using an effective Hamiltonian~\cite{sato96}, while 
Ellis and Tang\cite{ellis97} used chiral perturbation theory. 
Quark models have also been used to describe $\pi N$ 
scattering~\cite{vjt86,pa86,dspr95}.
Fuda~\cite{pncre1} developed a Poincar\'{e} invariant front form
model of $\pi N$ scattering, in which the potentials were  
assumed to be seperable. This model was later extended~\cite{pncre2} to
a pion laboratory kinetic energy of 1 GeV, giving an excellent
description of the phase shifts and inelasticities in the
$s$-, $p$- and $d$-waves.

The exact $\pi N \leftarrow \pi N$ amplitude for a given Lagrangian
can in principle be obtained from the full BS equation, with fully dressed 
propagators in the $\pi N$ intermediate state, and a potential consisting of 
all 1- and 2-particle irreducible connected diagrams.
By definition this exact amplitude would satisfy both crossing and chiral symmetry,
and have the correct one-body limit (an equation is said to have the 
one-body limit if it correctly reduces to either the Klein-Gordon or Dirac equation
if the mass of either of the particles becomes infinitely heavy).
Unfortunately, at present it is impossible to construct the potential for
the full BS equation, as it would contain an infinite number of Feynman diagrams. 
Consequently the potential and propagators must be approximated in some way.
The simplest and most commonly used approximation involves replacing the dressed 
propagators in the intermediate states with bare propagators with poles at the 
physical masses, and truncating the potential so that it contains only the 
lowest-order diagrams, i.e. the tree-level diagrams.
The resulting equation is generally referred to as the ladder BS equation.
There are some problems with this equation, such as it does not 
have the correct one-body limit~\cite{grossobl}.
Also some symmetries present in the approximate potential,
such as crossing and chiral symmetry, 
may be violated in the solution of the BS equation~\cite{cjgt87}.

The BS equation for the $\pi N$ scattering amplitude is a covariant four-dimensional 
integral equation, in which the integration is over the relative energy and relative 
momentum of the $\pi N$ intermediate state. It is the presence of the relative 
energy as an integration variable that is responsible for the complicated singularity
structure of the kernel. 
Frequently the BS equation is reduced to a 3-D integral equation in order
to avoid the difficulties involved in the handling of the singularities of the kernel. 
This is achieved by approximating the kernel in such a way that the integration over the
relative energy can be carried out explicitly, resulting in a 3-D integral
equation.
There are an infinite number of 3-D reductions of the 
BS equation~\cite{yaes71}, all of which satisfy relativistic elastic 
two-body unitarity.
There is no overwhelming reason to choose one particular 3-D approximation over 
any other. Some of the 3-D equations are chosen so as to overcome problems caused 
by approximating the full BS equation with the ladder approximation. For example, 
the ``smooth propagator''~\cite{coop88}, used by  Pearce and Jennings~\cite{pj91} and
J\"{a}de~\cite{jade}, attempts to restore chiral symmetry. However, 
other problems are introduced
when the dimensionality is reduced from 4-D to 3-D. 
Pascalutsa and Tjon~\cite{tjon98} 
have shown that when the nucleon self-energy
is calculated using most of the commonly used 3-D propagators, there are differences
in the renormalization between the positive and negative energy states,
which is an indication that
charge conjugation and CPT symmetries are violated. Another 
problem is that there are significant differences in the half-off-shell
amplitudes as calculated by different 3-D reductions of the BS
equation~\cite{leeyang94}, even when they all give the same results on-shell.
This could have significant implications, in that 
it is the off-shell behaviour that is important when the $\pi N$ amplitude
is used as input into the calculation of other nuclear reactions, such as 
pion-nucleus scattering, pion production in nucleon-nucleon collisions, 
or pion photoproduction.

In the present investigation we describe a relativistic model of
 $\pi N$ scattering
in which the BS equation is solved directly in four-dimensions.
In this way we can avoid the ambiguities encountered in reducing the
BS equation to three-dimensions. This we hope will give us coupling 
constants that could be compared to those extracted from QCD models,
as it has been shown that there are considerable differences in
coupling constants obtained using 3-D equations as compared to the
BS equation~\cite{lahaf97}.
More important is the fact that our $\pi N$
amplitude can be gauged~\cite{aa95} to give a photoproduction amplitude
that satisfies unitarity and gauge invariance.

To our knowledge
there are currently no models based on the BS equation
that give the $s$- and $p$-wave $\pi N$ phase-shifts in good agreement with
the empirical data.
The only previous meson-exchange model for $\pi N$ scattering to
have used the 4-dimensional BS equation was that of 
Nieland and Tjon~\cite{tpinbse}.
Since the potential consisted only of the $u$-channel nucleon pole diagram, 
this model could only give the $P_{33}$ phase-shifts in agreement with experiment.
In order to get a good description of all the $s$- and $p$-wave phase shifts,
it is necessary to include in the potential $s$- and $u$-channel $N$ and $\Delta (1232)$
poles, and in addition $t$-channel $\rho$  and $\sigma$ exchange diagrams.
All of the recent $\pi N$ meson-exchange models have included these diagrams,
although Sch{\"u}tz {\it et al.} replaced the $\rho$  and $\sigma$ exchange diagrams with
correlated two-pion exchange~\cite{schutz94}. The model of Sch{\"u}tz 
was later extended~\cite{schutz98}
to include the coupling to the $\pi \Delta$, $\eta N$ and $\sigma N$ channels,
and included the $N^*(1535)$ pole diagrams in the potential. Pascalutsa and Tjon 
included the Roper resonance~\cite{tjon98}, and later also included the 
$S_{11}$ and $D_{13}$ resonances as elementary particle poles in the 
potential~\cite{tjon98a}. 
We do not include the Roper or any other
higher baryon resonances, or coupling to channels other than $\pi N$,
as these contributions are expected to be small for elastic $\pi N$ scattering and 
at pion energies below the Roper resonance.
Here we are mainly interested in $\pi N$ scattering below the two-pion 
production threshold (around 360 MeV pion laboratory energy).

The organization of this paper is as follows.
In Sec.~\ref{sec.2} we state the Bethe-Salpeter equation and give
expressions for the dressed vertices and propagators. The Lagrangian
and form factors used in our model, and the choice of spin-3/2
propagator for the $\Delta (1232)$, are discussed in Sec.~\ref{sec.3}.
The renormalization procedure is outlined in Sec.~\ref{sec.4},
and we then proceed to describe our method of solving the BS
equation in Sec.~\ref{sec.5}. Here we also discuss the analytic
structure of the BS equation.
In Sec.~\ref{sec.7} we present results of fits to the
SM95 partial wave analysis of Arndt {\it et al.}~\cite{sm95}
from threshold up to 360 MeV pion laboratory energy. We compare the
coupling constants obtained in the present work to those
obtained using models of $\pi N$ scattering based on three-dimensional
reductions of the BS equation. We also discuss how the phase
shifts are built up from the individual diagrams in the potential, and
calculate the renormalized pion cutoff mass in order to examine the effect
of dressing on the $\pi N N$ form factor. Next we consider a second
model which differs from the first in the parameterization of
the form factors. We conclude this section by presenting results for the 
phase shifts up to 600 MeV.
Finally, in Sec.~\ref{sec.8} we present some concluding remarks.

\section{THE BS EQUATION FOR $\pi N$ SCATTERING}\label{sec.2}

We consider the scattering process
\beq 
\pi (p'_{\pi}) + N (p'_N) \longleftarrow \pi (p_{\pi}) + N (p_N) \, ,
\eeq
where $p_{\pi}$ and $p_N$ represent the incoming 4-momenta of the pion and
nucleon, while $p'_{\pi}$ and $p'_N$ are the outgoing 4-momenta.
The total 4-momentum $P$ is given by
\beq
P = p_{\pi} + p_N = p'_{\pi} + p'_N \, ,
\eeq
and the relative 4-momenta in the initial and final states are defined as
\beq
q=\half (p_N - p_{\pi}) \, , \hspace{1.0cm} 
q'=\half (p'_N - p'_{\pi}) \, .
\eeq
In the centre-of-mass (CM) frame the total 4-momentum is
related to the total energy $\sqrt{s}$ by $P=(\sqrt{s},\bold{0})$.
In addition, the Mandelstam variables $u$ and $t$ are given in terms
of the relative momenta as
\beq
u  =  (q + q')^2 \, ,\hspace{1.0cm} 
t  =  (q - q')^2 \, .
\eeq

Having defined the kinematics, 
the Bethe-Salpeter equation~\cite{bse} for 
the $\pi N \leftarrow \pi N$ amplitude $T(q',q;P)$ can be written as
\beq
T(q',q;P) = V(q',q;P) - {i \over (2 \pi)^4} \int d^4 q'' \, V(q',q'';P)
G_{\pi N} (q'';P) T(q'',q;P) \, , \label{eq:bse1}
\eeq
where $V$ is the potential.
The two-body $\pi N$ propagator $G_{\pi N}$ is the
product of the pion and nucleon propagators, i.e.
\beq
G_{\pi N}(q;P) = {1 \over (P/2-q)^2 - m_{\pi}^2+i \epsilon} \,
{\sla{P}/2+ \slas{q}+ m_N \over 
(P/2+q)^2 - m_N^2 +i \epsilon} \, , \label{eq:green1}
\eeq
where $m_{\pi}$ and $m_N$ 
are the charge-averaged pion and nucleon masses. 
In principle both the nucleon and pion propagators in
the $\pi N$ intermediate states should be dressed. However,
since here we are only requiring that two-body unitarity be
maintained, we have replaced the
dressed nucleon and pion propagators with bare propagators 
with poles at the physical nucleon and pion masses.

\begin{figure} 
\begin{center}
\epsfig{figure=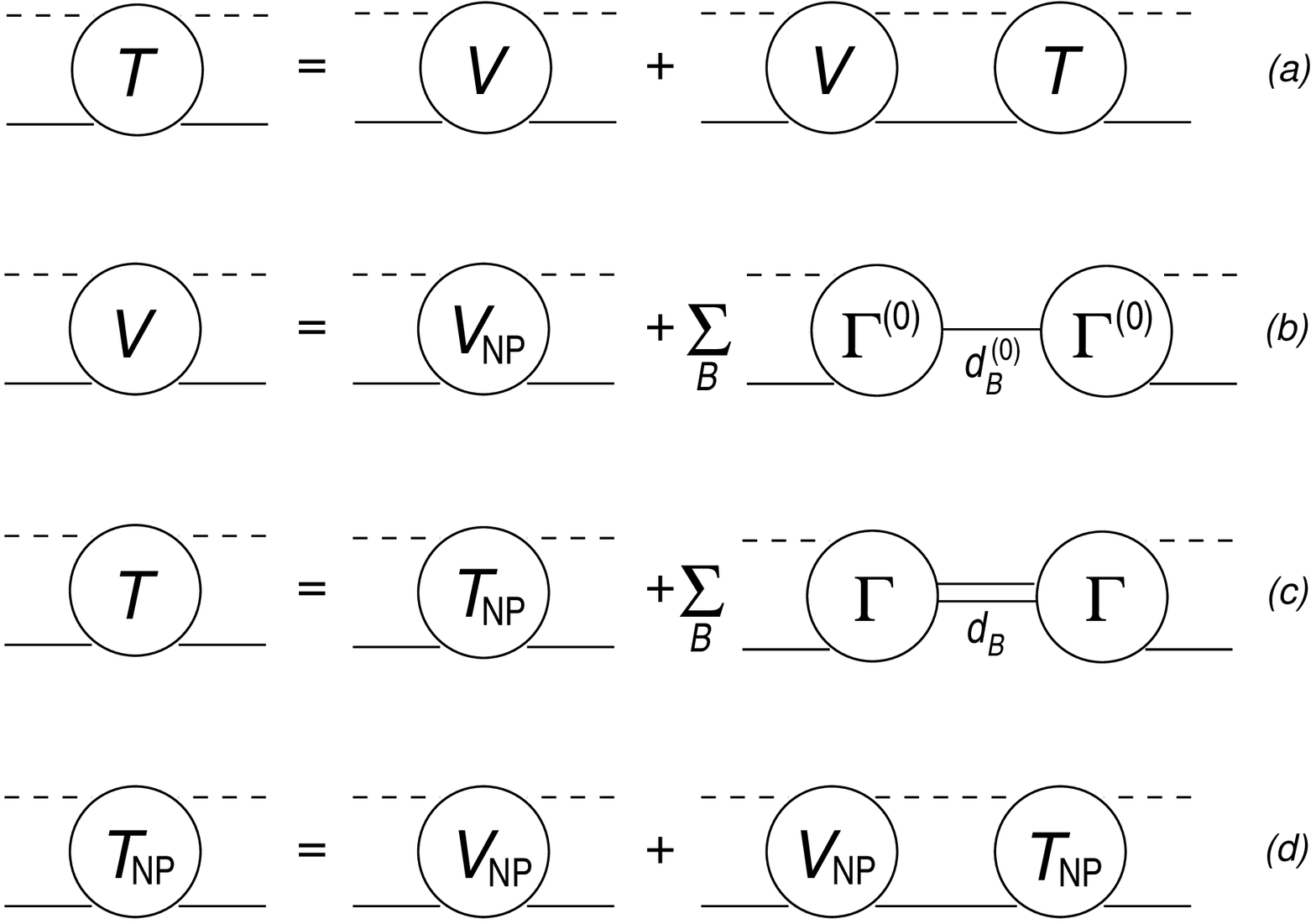,width=10.5cm,scale=1.0,angle=-90}
\end{center}
\caption{Graphical representation of: (a) the BS equation
for the full $T$-matrix, (b) the potential and (c) the $T$-matrix in terms
of non-pole and pole parts, and (d) the BS equation for the non-pole
$T$-matrix.}
\label{fig:bseqs}
\end{figure}

The potential $V(q',q;P)$ is constructed from the sum of 
$s$-, $t$- and $u$-channel pole diagrams (see Sec.~\ref{sec.3}).
It is well known that when an $s$-channel pole is included
in the potential of the ladder BS equation, or one of its 3-D
approximations, the solution of the BS equation also contains an 
$s$-channel pole diagram, in which the propagator and vertices are 
dressed~\cite{hay69}. 
The potential can be divided into the sum of non-pole and pole
contributions,
\beq
V(q',q;P) = V_{\mbox{\scriptsize NP}}(q',q;P) + 
\sum_{B} \Gamma ^{(0) \dagger}_{\pi N B}(q';P) \,
d^{(0)}_{B}(P) \, \Gamma ^{(0)}_{\pi N B}(q;P) \label{eq:bsep1} \, ,
\eeq
where $\Gamma ^{(0)}_{\pi N B}$ is the bare $\pi N B$ vertex, and
$d^{(0)}_{B}$ is the bare propagator for baryon $B$. At present the
only baryons we include in the potential are the nucleon and $\Delta (1232)$,
and so we have $B=N$, $\Delta$.
The pole part of Eq.~(\ref{eq:bsep1}) consists of the sum of $s$-channel baryon pole diagrams 
while $V_{\mbox{\scriptsize NP}}$, the non-pole part of the
potential, contains the $u$- and $t$-channel exchange diagrams. 
With the potential having this form, the solution of the BS equation can be written 
in a similar way as 
\beq
T(q',q;P) = T_{\mbox{\scriptsize NP}}(q',q;P) + 
\sum_{B} \Gamma ^{\dagger}_{\pi N B}(q';P) \,
d_{B}(P) \, \Gamma _{\pi N B}(q;P) \label{eq:bsep2} \, ,
\eeq
where $\Gamma _{\pi N B}$ and $d_{B}$ are the dressed $\pi N B$ vertex and dressed
baryon propagator respectively.
The non-pole part of the $T$-matrix
$T_{\mbox{\scriptsize NP}}$ is the solution of the BS equation
with a potential consisting of the sum of the $u$- and $t$-channel poles, i.e.
\beq
T_{\mbox{\scriptsize NP}}(q',q;P) = 
V_{\mbox{\scriptsize NP}}(q',q;P) - 
{i \over (2 \pi)^4} \int d^4 q'' \, V_{\mbox{\scriptsize NP}}(q',q'';P)
G_{\pi N} (q'';P) T_{\mbox{\scriptsize NP}}(q'',q;P) \, . \label{eq:bse2}
\eeq
Equations (\ref{eq:bse1}), (\ref{eq:bsep1}), (\ref{eq:bsep2}) and (\ref{eq:bse2}) are
shown diagrammatically in Fig.~\ref{fig:bseqs}.

\begin{figure}
\begin{center}
\epsfig{figure=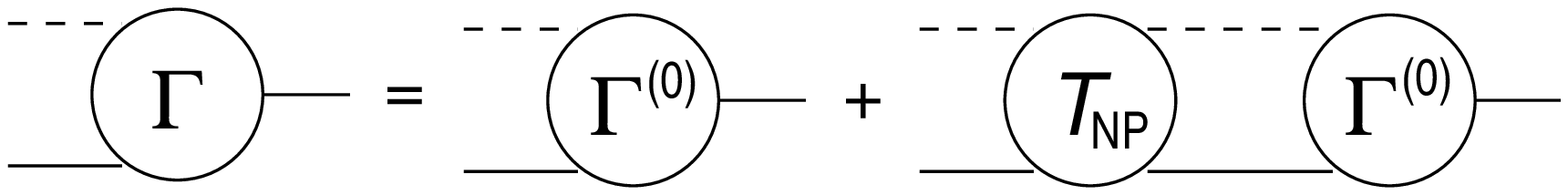,width=1.5cm,scale=1.0,angle=-90}
\end{center}
\caption{The equation for the dressed $\pi N B$ vertex.}
\label{fig:drvert}
\end{figure}

\begin{figure}
\begin{center}
\epsfig{figure=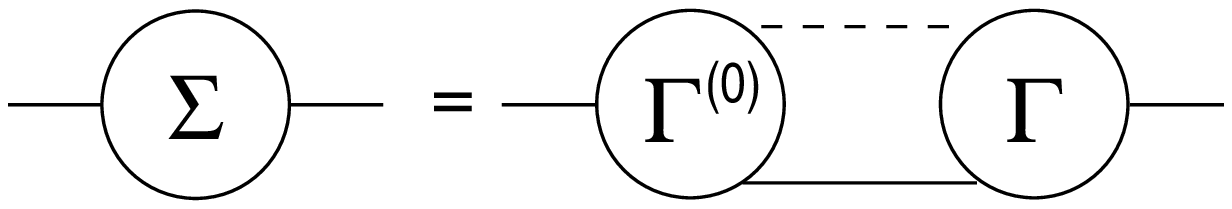,width=1.5cm,scale=0.5,angle=-90}
\end{center}
\caption{The baryon self-energy.}
\label{fig:drnucl}
\end{figure}

The dressed $\pi N N$ vertex is given in terms of the bare vertex 
and non-pole part of the $T$-matrix as (see Fig.~\ref{fig:drvert})
\beq
\Gamma_{\pi N N} (q;P) = \Gamma ^{(0)}_{\pi N N} (q;P) 
- {i \over (2 \pi)^4}\int d^4 q'' \, 
\Gamma _{\pi N N}^{(0)}  (q'';P)G_{\pi N}(q'';P) 
T_{\mbox{\scriptsize NP}}(q'',q;P) \label{eq:ndressv} \, .
\eeq
The bare and dressed nucleon propagators are
\bea
d_N^{(0)}(P) & = & \left[ \sla{P} - m_N^{(0)} + i \epsilon \right]^{-1} \, , \\
d_N (P) & = & \left[ \sla{P} - m_N^{(0)} - \Sigma _N (P)
+ i \epsilon \right]^{-1} \, ,
\eea
where $m_N^{(0)}$ is the bare nucleon mass.
Also, the nucleon self-energy $\Sigma _N (P)$ is  given by
\beq
-i \Sigma _N (P) = - {1 \over (2 \pi)^4} 
\int d^4 q \, \Gamma ^{(0)}_{\pi N N} (q;P)
G_{\pi N}(q;P) \Gamma ^{\dagger }_{\pi N N}(q;P) 
\, , \label{eq:nselfe}
\eeq
and is illustrated in Fig.~\ref{fig:drnucl}. Making use of 
Eq.~(\ref{eq:ndressv}), the nucleon self-energy can be
written as
\bea
-i \Sigma _N (P) &=& - {1 \over (2 \pi)^4} 
\int d^4 q \, \Gamma ^{(0)}_{\pi N N} (q;P)
G_{\pi N}(q;P) \Gamma ^{(0) \dagger }_{\pi N N}(q;P) \nonumber \\
& & \hspace{-1.5cm}
+ {i \over (2 \pi)^8} \int d^4 q' d^4 q \, \Gamma ^{(0)}_{\pi N N} (q';P)
G_{\pi N}(q';P) T_{\mbox{\scriptsize NP}} (q',q;P) G_{\pi N}(q;P)
 \Gamma ^{(0) \dagger }_{\pi N N}(q;P) \, . \label{eq:nselfen}
\eea
The first term in the right-hand-side of
Eq.~(\ref{eq:nselfen}) corresponds to the one-pion
loop dressing of the nucleon, while the second term is the
contribution arising from the iteration of the non-pole part of 
the potential in the BS equation (we refer to this term as the non-pole
contribution to the self-energy).
Since in the $\pi N$ intermediate states
we have approximated the dressed nucleon propagator by a
bare nucleon propagator with a pole at the physical nucleon mass,
we avoid having to solve a non-linear Schwinger-Dyson equation 
for the nucleon self-energy.

There are similar expressions to Eqs.~(\ref{eq:ndressv}) and
(\ref{eq:nselfe}) for the dressed $\pi N \Delta$
vertex and the $\Delta$ self-energy, however we do not calculate
them explicitly, for reasons given in Sec.~\ref{sec.4}.

\section{THE POTENTIAL}\label{sec.3}

The potential, or driving term,
which is iterated to all orders in the BS
equation to obtain an amplitude satisfying two-body
unitarity,
consists of all the tree-level Feynman diagrams contributing
to the process $\pi N \leftarrow \pi N$, derived from the
interaction Lagrangian under consideration. In this section we discuss 
the Lagrangian used, the coupling constants in this Lagrangian,
and the choice of form factors which are necessary to obtain convergence.
We also look at the possible different forms of spin-3/2 propagators 
that can be used for the $\Delta$.

\subsection{The Lagrangian}

\begin{figure}
\begin{center}
\epsfig{figure=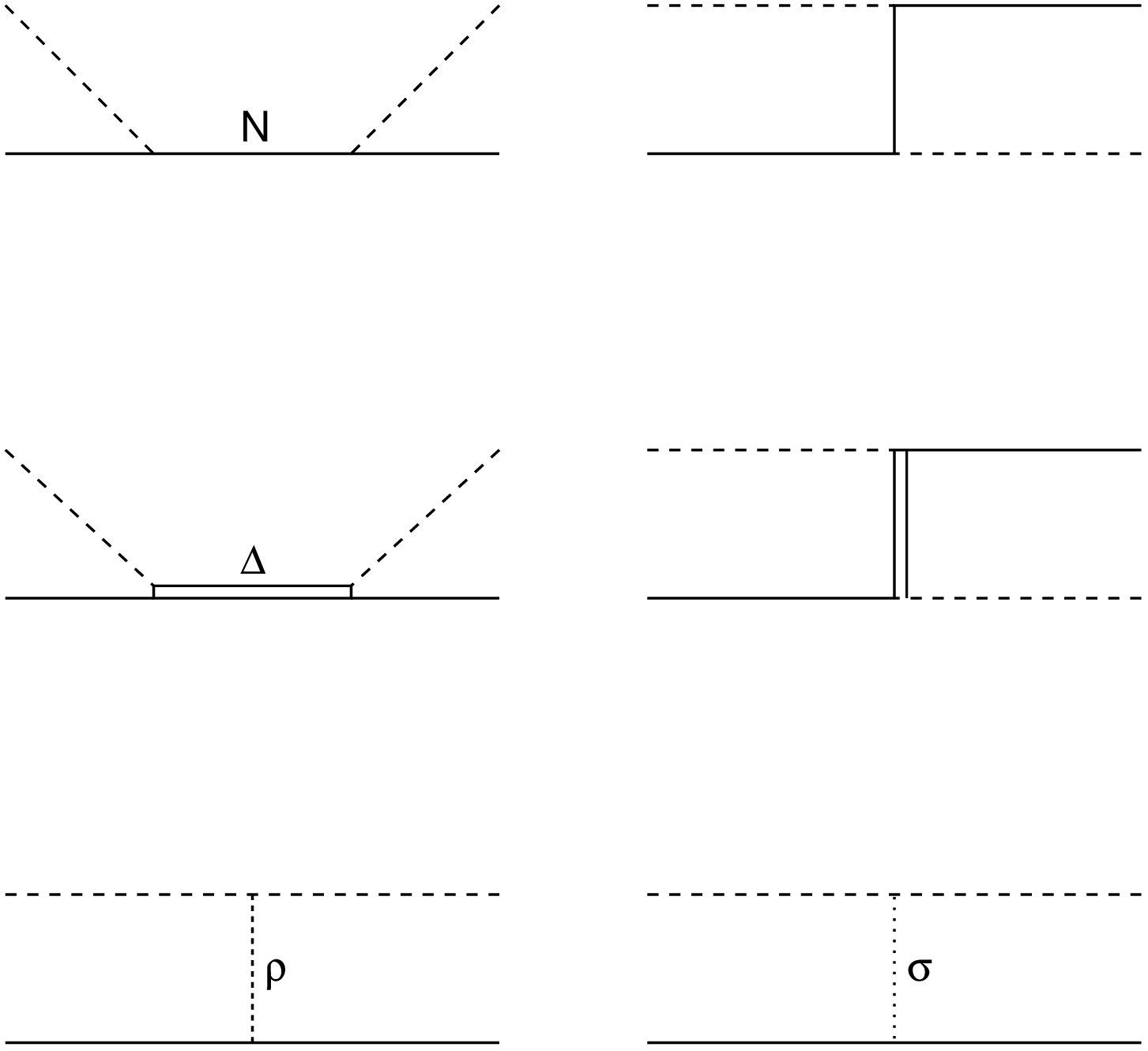,width=10.5cm,scale=0.5,angle=-90}
\end{center}
\vspace*{0.5cm}
\caption{The diagrams included in the potential of the BS equation.}
\label{fig:fullpot1}
\end{figure}

The tree-level diagrams shown in Fig.~\ref{fig:fullpot1} are obtained
from the following interaction Lagrangian:
\bea
\lag _{\mbox{\scriptsize int}} &=& {g_{\pi N N} \over 2 m_N} \,
\bar \psi _N \, \gamma _5 \, \gamma ^{\mu} \, \bbox{\tau} \cdot 
\partial _{\mu} { \bbox{\pi}} \, \psi _N
+ {f_{\pi N \Delta} \over \mpi} \, 
\bar \psi ^{\mu}_{\Delta} \,\left( g_{\mu \nu} + x_{\Delta} \gamma _{\mu}
\gamma _{\nu} \right) \bold{T}  \psi _N 
\cdot \partial ^{\nu} \bbox{\pi}
 + \mbox{h.c.} \nonumber \\
& & +g_{\rho N N} \bar \psi _N \half \bbox{\tau} \cdot
\left( \gamma ^{\mu} \bbox{\rho }_{\mu} +
{\kappa _{\rho} \over 2 m_N} \sigma^{\mu \nu} \partial _{\mu}
\bbox{\rho} _{\nu} \right) \psi _N + g_{\rho\pi\pi}{\bbox{\rho}} 
^{\mu}\cdot (\partial _{\mu}
{ \bbox{\pi}} \times {\bbox{\pi}}) \nonumber \\
& & + g_{\sigma N N} \bar \psi _N \psi _N \sigma
 + {g_{\sigma \pi\pi} \over 2 \mpi} \,
\sigma \, \partial _{\mu} \bbox{\pi} \cdot \partial ^{\mu}\bbox{\pi}  \, , \label{eq:lag}
\eea
where $\psi _N$, $\psi ^{\mu} _{\Delta}$, $\bbox{\pi}$, $\bbox{\rho} _{\mu}$,
and $\sigma$ are the fields for the nucleon, delta, pion, rho and sigma,
respectively.
The derivative couplings of the pion field to the other mesons and
baryons ensures that chiral symmetry is satisfied at tree-level.
In the $\pi N \Delta$ Lagrangian,
$\bold{T}$ is the transition operator between isospin-3/2 and 1/2 states,
and $x_{\Delta}$ is a parameter that can be adjusted, and its value will be
considered when we discuss our choice for the $\Delta$ propagator.

Ideally, all of the coupling constants should be fixed using information 
from other sources, rather than leaving them as free parameters.
Out of all of the coupling constants appearing in the Lagrangian given in
Eq.~(\ref{eq:lag}), the $\pi NN$ coupling is the best known
(although even the $\pi N N$ coupling constant is not without 
controversy~\cite{ericson,mcmr}). We use the value advocated
by the Nijmegen group~\cite{pinn}, i.e. $g_{\pi N N}^2 /4 \pi = 13.5$.
The remaining coupling constants can be determined in a variety of
different ways, but there are some discrepancies.

A value for the
$\pi N \Delta$ coupling constant can be obtained by calculating the width
for the decay $\Delta \rightarrow \pi N$. The coupling constant is chosen
such that the width is equal to its experimental value.
Assuming that the $\Delta$
self-energy is dominated by the one-pion loop diagram, it is found that
$f^2_{\pi N \Delta}/ 4 \pi = 0.36$ gives the correct width. This result
makes use of the width calculated on the real $s$-axis, however it has
been shown that there are differences between widths calculated on the
real axis and in the complex plane~\cite{elsey}. A somewhat smaller value 
for $f_{\pi N \Delta}$ is found using the quark-model relation~\cite{bw75}
\beq
{f_{\pi N \Delta}^2 } = {72 \over 25}
\left({m_{\pi} \over 2 m_N} \right) ^2 {g^2_{\pi N N} } \, , \label{eq:dqmr}
\eeq
which gives  $f^2_{\pi N \Delta}/ 4 \pi = 0.21$ when the Nijmegen value
is used for $g_{\pi N N}$. This smaller value has also been shown to be
consistent with the width of the $\Delta$~\cite{nisk},
provided that higher-order mesonic corrections to the $\Delta$ self-energy 
are included along with the one-pion loop term.

We now consider the $\rho$ exchange diagram, which depends only on the
product $g^2_{\rho} \equiv g_{\rho \pi \pi} g_{\rho N N}$, as well as 
$\kappa _{\rho}$, the ratio of the tensor to vector $\rho N N$ coupling
constants.
There are a number of different ways of determining a value for $g_{\rho}$,
as originally discussed by Sakurai~\cite{sakurai}.
If it is assumed that the isospin-odd 
$\pi N$ scattering length is dominated by the tree-level
$\rho$ exchange diagram, then $g^2_{\rho}/4 \pi = 3.1$ is obtained.
Alternatively, it can be
assumed that the $\rho$ meson couples to both pions and nucleons with the
same strength (universality), which means that
$g_{\rho} = g_{\rho \pi \pi} = g_{\rho N N}$.
An estimation of $g_{\rho \pi \pi}$ can be obtained from the decay
$\rho \rightarrow  2 \pi$, which gives $g^2_{\rho \pi \pi}/ 4 \pi = 2.84$. A
very similar value of $g_{\rho \pi \pi}^2/4 \pi = 2.7$
is given by the
Kawarabayashi-Suzuki-Riazuddin-Fayyazuddin (KSRF) relation~\cite{ksrf},
which is obtained from current algebra and PCAC,
and states that $m_{\rho}^2 = 2 g^2_{\rho \pi \pi} f_{\pi}^2$, where
$f_{\pi}=93$ MeV is the pion decay constant. 
There is some controversy about the value of $\kappa _{\rho}$. The vector meson
dominance model (VMD) gives $\kappa _{\rho} = 3.7$~\cite{peccei}, while 
H\"{o}hler and Pietarinen found $\kappa _{\rho}= 6.6$ from a
$\pi \pi - N \bar N$ partial wave analysis~\cite{hohpie}.

Although the scalar $\sigma$ meson does not seem to exist in nature,
$t$-channel $\sigma$ exchange is included in $\pi N$ models as an
effective interaction, representing higher-order processes not explicitly
included in the potential, such as correlated $2 \pi$-exchange in the scalar-isoscalar channel. There
are therefore no reliable determinations of the mass of the $\sigma$ meson or
the magnitudes (or signs) of the $\sigma N N$ and $\sigma \pi \pi$ coupling 
constants. The mass is usually taken to be around $4m_{\pi}$ to $6m_{\pi}$.

\subsection{The $\Delta$ propagator}

There is an ambiguity as to the choice of propagator for a particle
with spin-3/2, and a number of different propagators have been 
introduced~\cite{rs,will85,pasv98,haberz98}. One way of deriving a 
spin-3/2 propagator is to
begin with the free Lagrangian for a massive spin-3/2 field~\cite{nath}. The
propagator obtained from this Lagrangian has, in its most general
form, a pole part and a non-pole part. The pole part is unique while
the non-pole part depends on a complex parameter $A$. When the form
of the $\pi N \Delta$ vertex is chosen correctly, the $S$-matrix and physical
quantities are independent of $A$~\cite{krs61}. Taking $A=-1$ gives the
simplest form for the propagator, and is commonly called the
Rarita-Schwinger (RS) propagator:
\beq
P_{\mu\nu}(p) = {\slas{p} + m_{\Delta} \over
p^2 - m_{\Delta}^2 + i \epsilon}  
\left[g_{\mu\nu} - {1 \over 3} \, \gamma _{\mu}
\gamma _{\nu} 
 -{1 \over 3 m_{\Delta}}
\, (\gamma _{\mu} p _{\nu} - \gamma _{\nu} p _{\mu}) -
{2 \over 3 m_{\Delta}^2} \, p_{\mu} p_{\nu} \right] \, .
\eeq
It is known that the RS propagator contains 
background or off-mass-shell spin-1/2 components, 
along with the spin-3/2 component (see, e.g. Ref.~\cite{ben89}).
This becomes evident when the 
RS propagator is written in terms of spin projection operators, which are 
denoted by ${\cal{P}}^J_{i j}$, and are given by~\cite{pvann,ben89}
\bea
({\cal{P}}^{3/2})_{\alpha \beta} & = & g_{\alpha \beta} - {1 \over 3}\gamma _{\alpha}
\gamma _{\beta} - {1 \over 3 p^2} \left(\slas{p} \gamma _{\alpha}
p_{\beta} + p _{\alpha} \gamma _{\beta} \slas{p} \right)  \, , \\
({\cal{P}}_{11}^{1/2})_{\alpha \beta} & = & {1 \over 3}\gamma _{\alpha}
\gamma _{\beta} - {p _{\alpha} p_{\beta} \over p^2} + {1 \over 3 p^2}
\left(\slas{p} \gamma _{\alpha}
p_{\beta} + p _{\alpha} \gamma _{\beta} \slas{p} \right) \, ,  \\
({\cal{P}}_{22}^{1/2})_{\alpha \beta} & = & {p _{\alpha} p_{\beta} \over p^2} \, ,  \\
({\cal{P}}_{21}^{1/2})_{\alpha \beta} & = & {1 \over \sqrt{3} p^2} 
\left( p _{\alpha} p _{\beta} - \slas{p} 
\gamma _{\alpha} p _{\beta} \right) \, , \\
({\cal{P}}_{12}^{1/2})_{\alpha \beta} & = & {1 \over \sqrt{3} p^2} 
\left( \slas{p} p _{\alpha} 
\gamma _{\beta} - p _{\alpha} p _{\beta} \right)  \, .
\eea
The RS propagator can then be written in terms of the spin
projection operators defined above, as
\beq
P_{\mu\nu}(p)  =  {\slas{p} + m_{\Delta} \over
p^2 - m_{\Delta}^2 + i \epsilon} {\cal{P}}^{3/2} - {2 \over 3 m_{\Delta}^2 }
\left(\slas{p} + m _{\Delta} \right) {\cal{P}}_{22}^{1/2} 
 + {1 \over \sqrt{3} m_{\Delta}} 
 \left( {\cal{P}}_{12}^{1/2} + {\cal{P}}_{21}^{1/2} \right) \, .
\eeq
The spin-1/2 background 
could be considered as being unphysical, since the $\Delta (1232)$ is known 
experimentally to be a particle with spin-3/2.

Williams introduced a propagator proportional to the
spin-3/2 projection operator~\cite{will85},
\beq
P_{\mu\nu}(p) = { \slas{p} + m_{\Delta} \over
p^2 - m_{\Delta}^2 + i \epsilon} \, {\cal{P}}^{3/2} \, .
\eeq
The Williams and RS propagators are identical when the $\Delta$ is
on-mass-shell, since the spin-1/2 components in the RS propagator are
only present when the $\Delta$ is off-mass-shell. Also, when the
Williams propagator is used there are no contributions to the 
$\pi N \leftarrow \pi N$ amplitude arising from the parts of the
$\pi N \Delta$ vertices proportional to $x_{\Delta}$.

There have been attempts to fix $x_{\Delta}$ on theoretical grounds.
Peccei~\cite{peccei} suggested that the choice $x_{\Delta} = -1/4$
ensures that there is no direct coupling to the spin-1/2 components
of the RS propagator, but it was later shown~\cite{ben89}
that the spin-1/2 components are always present, and cannot
be removed by choosing a particular value of $x_{\Delta}$.
Nath {\it et al.}~\cite{nath} suggested that $x_{\Delta}=-1$ should be used
if the $\pi N \Delta$ vertex is to
be consistent with the principles of second quantization.

The $1/p^2$ factor in the Williams propagator can cause numerical 
difficulties~\cite{bof91,gross93}. Pascalutsa~\cite{pasv98} used 
the Hamiltonian
path-integral formulation to investigate the interacting spin-3/2
field and constructed a theory in which there is no 
coupling to the spin-1/2 components. For the case of the
$s$- and $u$-channel tree-level amplitudes for $\pi N$ scattering,
the $\pi N \Delta$ vertex corresponds to the usual $\pi N \Delta$
vertex with $x_{\Delta}=0$, and the $\Delta$ propagator is the same
as the Williams propagator but multiplied by 
$p^2/ m_{\Delta}^2$. This extra factor of $p^2$ in the numerator fixes the
problems caused by the $1/p^2$ term in the spin-3/2 projection 
operator.

We consider two possibilities for the $\Delta$ propagator: (i)
the Rarita-Schwinger propagator and general $\pi N \Delta$
vertex, with $x_{\Delta}$ as a free parameter, and (ii) the Pascalutsa 
propagator and vertex.
In both cases the $\Delta$ is treated as a stable particle in the
potential, and the width is generated dynamically, since the bare $s$-channel 
$\Delta$ pole diagram is dressed when the BS equation is solved.

\subsection{Form factors}

The integrals in 
Eqs.~(\ref{eq:bse1}), (\ref{eq:bse2}), (\ref{eq:ndressv}), and (\ref{eq:nselfen})
are divergent, and so a regularization scheme must be
implemented in order to obtain finite results. As is commonly done in
meson-exchange models, we introduce
form factors at each interaction vertex. These  form factors 
represent the extended structure of the
particles involved, and will  ensure that all integrals are convergent
by suppressing contributions at high momenta.
Since at present it is not possible to  calculate the appropriate form
factors directly from QCD, phenomenological functions are usually
chosen as the form factors, which have no connection to the underlying
quark dynamics.
A consequence of this is that free parameters, the cutoff 
masses that govern the range of suppression,
are introduced into the model.

In meson-exchange models of the $NN$ interaction, the form factors depend only on the
4-momentum squared of the exchange particles. This cannot be done in
$\pi N$ models because, for example, the form factors would provide no
convergence for the $s$-channel pole diagrams. We therefore follow
Pearce and Jennings~\cite{pj91}, and make the assumption that the cutoff
function associated with each vertex is a product of form factors
that depend on the 4-momentum squared of each particle present at
the vertex. The $a b c$ vertex is therefore given by
\beq \Gamma _{a b c} = f_{a b c}(q^2_{a},q^2_{b},q^2_{c}) \, {\cal V} _{a b c} \, ,
\eeq
where ${\cal V} _{a b c}$ is the coupling operator obtained
from the interaction Lagrangian, Eq.~(\ref{eq:lag}),
and the associated form factor is of the separable form:
\beq
f_{a b c}(q^2_{a},q^2_{b},q^2_{c}) =
f_{a}(q^2_{a}) \,
f_{b}(q^2_{b}) \, f_{c}(q^2_{c}) \label{eq:sepformdef} \, .
\eeq
The 4-momenta squared of the legs of the vertices
are denoted by $q_a^2$, $q_b^2$ and $q_c^2$.
It is conventional to  choose the normalization
such that $f(m^2)=1$, where $m$ is the mass of the corresponding
particle. Therefore, at the unphysical point 
when all three legs of a vertex are on-mass-shell,
the corresponding product of form factors is equal to one.

The scalar functions $f(q^2)$ can essentially be chosen in an {\it ad hoc} 
manner, since very little is known about the off-mass-shell behaviour
of the form factors. 
One possible choice is the multipole form factor:
\beq
f_{\mbox{\scriptsize I}}(q^2) = \left({ \Lambda ^2 - m^2
\over \Lambda ^2 - q^2} \right) ^{n} \, , \label{eq:formI}
\eeq
where $\Lambda $ is the cutoff mass, and $n$ is an integer.
Different forms of $f(q^2)$ have been used in the various models
of $\pi N$ scattering.
An example is~\cite{gross93,leeyang94}
\beq
f_{\mbox{\scriptsize II}}(q^2) = \left( {(\Lambda ^2-m^2)^2 \over 
(\Lambda ^2 - m^2)^2 + (m^2 - q^2)^2} \right) ^n \, . \label{eq:formII}
\eeq
The main difference between the two form factors 
$f_{\mbox{\scriptsize I}}$ and $f_{\mbox{\scriptsize II}}$
lies in the analytic structure in the complex $q_0$ plane 
(note $q^2=q_0^2 - \bold{q}^2 $).
The function $f_{\mbox{\scriptsize I}}$ has two poles along the real $q_0$ axis at
\beq
q_0 = \pm \left( \sqrt{\Lambda^2 + \bold{q}^2} - i \epsilon \right) \, ,
\eeq
while $f_{\mbox{\scriptsize II}}$
 has four poles in the complex $q_0$ plane, located at
\beq
q_0 = \pm \sqrt{\Lambda^2 \pm i (\Lambda^2-m^2) + \bold{q}^2} \, .
\eeq
The form factors used by Pearce and Jennings~\cite{pj91} have a 
very similar structure to $f_{\mbox{\scriptsize II}}$
in the complex $q_0$ plane.
We note that $f_{\mbox{\scriptsize II}}(q^2)$ always has poles in each
of the four quadrants of the $q_0$ complex plane, 
irrespective of the value of $\Lambda$ (except when
$\Lambda^2 = m^2$).  We should also mention that 
$f_{\mbox{\scriptsize I}}(q^2)$ with $n=1$ could be
considered as the propagator for a scalar particle with mass $\Lambda$,
while $f_{\mbox{\scriptsize II}}(q^2)$ with $n=1$
corresponds to the propagator for a resonance with
mass $\Lambda$, and a width proportional to
$(\Lambda^2-m^2)$. It has been shown that when a form factor
has the form of a propagator for a resonance with a constant width, 
there is a violation of unitarity at all energies, which only becomes evident 
when a four-dimensional formulation is used~\cite{pena98}. 

As an example of the problems caused by this violation of unitarity, we can
consider the dressed nucleon propagator. If the one-pion
loop self-energy diagram is calculated using form factors similar to
$f_{\mbox{\scriptsize II}}(q^2)$ at each vertex, it turns
out that the dressed nucleon has a width. This is of course unphysical, since
the nucleon is well known to be stable particle. As a result, 
form factors with poles in the complex $q_0$ plane cannot be used to
regularize loop diagrams in 4-D models. Therefore, here
we use form factors of the form $f_{\mbox{\scriptsize I}}$, in which case
the unitarity violations do not occur.

\section{RENORMALIZATION}\label{sec.4}

As shown in Sec.~\ref{sec.2},
the bare vertices and propagators appearing in the $s$-channel pole
diagrams in the potential become dressed when the potential 
is iterated in the BS equation.
A renormalization procedure therefore must be carried out in order to fix
the bare parameters such that the renormalized quantities are equal to
their physical values. Since we include form factors at all vertices, 
the bare masses and coupling constants are finite.

\subsection{The dressed $\pi NN$ vertex}

The $\pi N N$ vertex renormalization constant $Z_{1N}$ is defined in the 
usual way:
the bare $\pi N N$ vertex differs from the dressed vertex by only a constant,
which is $Z_{1N}$, when sandwiched between Dirac spinors and  all external
legs are placed on-mass-shell. Therefore
the vertex renormalization constant is defined by the relation
\begin{equation}
\bar u ( \bold{P} ) \Gamma _{\pi N N}(q;P) u (\bold{q}) = 
Z_{1N}^{-1} \bar u ( \bold{P} ) 
\Gamma _{\pi N N}^{(0)}(q;P) u (\bold{q}) \, , \label{eq:z1def}
\end{equation}
with $P^2=m_N^2$, and $q^2$ taken such that
all three legs of the vertex are on-mass-shell.

\subsection{The dressed nucleon propagator}

We now turn to the renormalization of the dressed nucleon
propagator.
Firstly, Lorentz invariance requires that the nucleon 
self-energy can be written as the sum of a vector and scalar part, i.e.
\beq
\Sigma _N (P) = \sla{P} A(s) + m_N B(s) \, ,
\eeq
where $A(s)$ and $B(s)$ are functions of $s$ only.
In order to fix the bare nucleon mass, we require that the dressed propagator has a 
pole at the physical nucleon mass, i.e.
\beq
\lim_{s \rightarrow m_N^2} d^{-1}_N(P) = 0 \, .
\eeq
This requirement gives the following expression
for the bare nucleon mass in terms of the
functions $A$ and $B$:
\beq
m_N^{(0)} = m_N \left( 1 - A(m_N^2) - B(m_N^2) \right) \label{eq:bnme} \, .
\eeq
The residue of the dressed nucleon propagator at the nucleon pole
is defined as
the nucleon wave-function renormalization constant $Z_{2N}$, i.e.
\beq
\lim_{s \rightarrow m_N^2} (\sla{P} - m_N) d_N(P) = Z_{2N} \, ,
\eeq
which gives~\cite{aa95}
\beq
Z_{2N} = \left[ 1 - A(m_N^2) - 2 m_N^2 \left(A'(m_N^2) + B'(m_N^2) \right)
\right] ^{-1} \label{eq:z2e} \, ,
\eeq
where
\beq	
A'(m_N^2) = {d \over ds}A(s) \Biggl| _{s=m_N^2} \, ,
\hskip 1cm
B'(m_N^2) = {d \over ds}B(s) \Biggl| _{s=m_N^2} \label{eq:apbp} \, .
\eeq
The above expressions enable the renormalization constants to be
calculated easily for the case of one-pion loop dressing,
since it is not hard in this case to write $\Sigma _N(P)$ in 
terms of the functions
$A(s)$ and $B(s)$. However, if the non-pole contributions to the
nucleon self-energy are taken into account, then this would not be
the simplest method of calculation.
We introduce the scalar quantity $\Sigma _N ^{\bar u u} (s)$, which is
defined as the nucleon self-energy sandwiched between Dirac spinors, i.e.
\bea
\Sigma ^{\bar u u}_N (s) &=& \bar u (\bold{P}) \Sigma _N (P) 
u (\bold{P}) \nonumber \\
& = & \sqrt{s} A(s) + m_N B(s) \, ,
\eea
since $\bold{P}=\bold{0}$ in the CM system.
Taking note of Eqs.~(\ref{eq:bnme}), (\ref{eq:z2e}), and (\ref{eq:apbp}),
the renormalization constants $m_N^{(0)}$ and 
$Z_{2N}$ can be written in terms 
of $\Sigma _N ^{\bar u u} (s)$ as
\bea
m_N^{(0)} & = & m_N - \Sigma _N ^{\bar u u} (m_N^2) \, , \label{eq:mnoc}\\
Z_{2N} &=& \left[ 1 - 2m_N {d \over ds }\Sigma _N ^{\bar u u} (s) \Biggl| _{s=m_N^2}
\right] ^{-1} \, . \label{eq:z2c}
\eea
In this way we can include both the one-loop and the non-pole contributions
to the mass shift and wavefunction
renormalization $Z_{2N}$ in a way consistent with the
scattering formulation of the BS equation.

\subsection{Renormalization of the BS equation}

In the $\pi N$ amplitude there should be factors of $\sqrt{Z_{2 N}}$ for each of the
external nucleon legs, which result from the application of 
LSZ reduction~\cite{lsz55} 
on the $\pi N \leftarrow \pi N$ Green's function. However, since we are not
performing any explicit LSZ reduction, and we are assuming that the
nucleon
propagators in the kernel of the BS equation are bare propagators with
physical masses, the factors of $\sqrt{Z_{2 N}}$ are not generated, and the 
$\pi N N$ and $\pi N \Delta$ couplings are effective coupling constants that should
be set equal to the physical coupling constants.
The only exception is with the $s$-channel pole diagrams. In this case, the solution 
of the BS equation generates new $s$-channel pole amplitudes in which the baryon
propagators and $\pi N B$ vertices are dressed. This implies that bare masses and
coupling constants should be used in the $s$-channel pole diagrams 
in the potential.

In order that the $\pi N$ amplitude has a pole at the physical nucleon
mass, and that the residue at this pole is equal to the square of the physical $\pi N N$
coupling constant, the bare nucleon mass must be fixed by
Eq.~(\ref{eq:mnoc}), and the bare $\pi N N$ coupling constant fixed using
\beq
g^R_{\pi N N} = Z_{1N}^{-1} \sqrt{Z_{2N}} g_{\pi N N}^{(0)} \, .
\eeq
Here $g^{(0)}_{\pi N N}$ is the bare $\pi N N$ coupling constant, and
$g^R_{\pi N N}$ is the renormalized coupling constant, which is set equal
to the ``experimental'' $\pi N N$ coupling constant by fixing the value
of $g^{(0)}_{\pi N N}$ correctly.

In principle, a similar renormalization procedure should be carried out for the
$\Delta$. However, the pole in the $T$-matrix 
corresponding to the dressed $\Delta$ occurs in the
complex $s$-plane, since the dressed $\Delta$ has a width.
Therefore, in order to
fix the bare $\pi N \Delta$ coupling constant, it would be necessary to 
analytically continue the BS equation into the complex $s$-plane~\cite{elsey}.
Rather than doing this,
here both $m_{\Delta}^{(0)}$ and $f_{\pi N \Delta}^{(0)}$ are treated as free
parameters. Since the $P_{33}$ partial wave is dominated by the $s$-channel
$\Delta$ pole diagram, the bare $\Delta$ parameters are essentially
fixed by the $P_{33}$ phase-shifts. The position at which the phase shifts
go through $90^{\circ}$ is determined by the bare $\Delta$ mass,
and the width of the resonance is related to the
bare $\pi N \Delta$ coupling constant.

\section{SOLVING THE BS EQUATION}\label{sec.5}

To calculate quantities such as phase shifts and
scattering lengths, we solve Eq.~(\ref{eq:bse1}), i.e. the
BS equation with the potential consisting the $s$- and $u$-channel $N$ and
$\Delta$ poles as well as $t$-channel $\rho$  and $\sigma$ exchange.
In addition, the dressed $\pi N N$ vertex and nucleon
self-energy need to be calculated so that the nucleon renormalization
procedure can be performed.

\subsection{Partial wave expansion}

The nucleon propagator present in the 
 $\pi N$ intermediate states can be separated into positive and negative
energy components~\cite{garc93},
which allows us to write $G_{\pi N}$ in terms of projection
operators: 
\beq 
G_{\pi N}(q;P) = G^{\bar u u}(q_0,q;s) \Lambda^{+}(\bold{q}) -
G^{\bar v v}(q_0,q;s) \Lambda^{-}(-\bold{q})\, , \label{eq:green}
\eeq
where $G^{\bar u u}$ and $G^{\bar v v}$ are given by
\bea
G^{\bar u u}(q_0,q;s) &=& {m_N \over E_q} \, 
{1 \over \sqrt{s}/2+q_0-E_q+i \epsilon}
\, {1 \over (\sqrt{s}/2-q_0)^2-\omega _q^2+i \epsilon} \, , \\
G^{\bar v v}(q_0,q;s) &=& {m_N \over E_q} \, 
{1 \over \sqrt{s}/2+q_0+E_q-i \epsilon}
\, {1 \over (\sqrt{s}/2-q_0)^2-\omega _q^2+i \epsilon} \, ,
\eea
with $E_q=\sqrt{\bold{q}^2+m_N^2}$ and 
$\omega _q=\sqrt{\bold{q}^2+m_{\pi}^2}$.
The positive and negative energy 
projection operators can be written in terms of Dirac
spinors  as
\bea
\Lambda^{+}(\bold{q}) &=& \sum _{r}u_r(\bold{q}) 
\bar u_r(\bold{q}) \, , \\
\Lambda^{-}(\bold{q}) &=& - \sum _{r}v_r(\bold{q}) 
\bar v_r(\bold{q}) \, .
\eea 
The normalization of these spinors is defined by
$u^{\dagger}_r(\bold{q}) u_r(\bold{q})  = 
v^{\dagger}_r(\bold{q}) v_r(\bold{q})  = 1$.

The expansion given above in Eq.~(\ref{eq:green}) is substituted into the BS equation.
Multiplying the BS equation from the left and right by Dirac spinors yields
a pair of coupled integral equations. If we introduce the notation
(suppressing the Dirac indices)
\bea
T^{ \bar u u}(q',q;P) &=&\bar u (\boldp{q}) T(q',q;P)u(\bold{q}) \, , \\
T^{ \bar v u}(q',q;P) &=&\bar v (-\boldp{q}) T(q',q;P)u(\bold{q}) \, ,
\eea
and similarly for the potential, the BS equation can be 
written as two coupled equations for $T^{ \bar u u}$ and
$T^{ \bar v u}$:
\bea
T^{ \bar w u}(q',q;P) &= &  V^{ \bar w u}(q',q;P) \nonumber \\
& & - {i \over (2 \pi)^4} \sum _{w''=u,v} 
\int d^4 q'' \, V^{ \bar w w''}(q',q'';P)
G^{\bar w'' w''}(q'';P)T^{ \bar w'' u}(q'',q;P) \, ,
\eea
with $w=u \mbox{, } v$.
There is a similar set of two coupled equations for the amplitudes
$T^{\bar u v}$ and $T^{\bar v v}$, which are required along with
$T^{ \bar u u}$ and $T^{ \bar v u}$ in the calculation of $m_N^{(0)}$ and 
$Z_{2N}$.

We now reduce the number of dimensions from four to two, 
by removing the angular dependence using a partial wave
expansion. Including spinor indices again, we can write each amplitude
in the form
\beq
A^{ \bar w' w}_{\lambda' \lambda}
(q',q;P) = \chi ^{\dagger} _{\lambda '} \, \tilde{A}^{ \bar w' w}
(q',q;P) \, \chi _{\lambda} \, ,
\eeq
where $w$ and $w'$ are either $u$ or $v$, and
$\chi_{\lambda}$ is a Pauli spinor. The amplitude $\tilde{A}^{ \bar w' w}$ 
can be expanded in terms of partial wave amplitudes 
$A^{\bar w' w}_{\ell \ell ' j}$ as
\beq
\tilde{A}^{\bar w' w}
(q'_0,\boldp{q};q_0,\bold{q};s) =  N(q',q) 
\sum_{\ell \ell ' j m} 
{\cal Y}_{\ell j m} (\boldh{q}) A^{\bar w' w}_{\ell \ell ' j}
(q'_0,q';q_0,q;s) {\cal Y}^{\dagger}_{\ell ' j m }(\boldhp{q}) \, ,
\eeq
where $q = | \bold{q} |$ , $q' = | \boldp{q} |$,
and $N(q',q) = - (2 \pi )^{4} (q' q)^{-1}$.
Also, the generalized Legendre polynomials are given by
\beq
{\cal Y}_{\ell j m} (\boldh{q}) = \sum _{m_{\ell} m_{s}}
(\ell \, m_{\ell} \, \half \, m_{s}  \mid  j \, m) 
Y_{\ell m_{\ell}}(\boldh{q})\chi _{m_s} \, ,
\eeq
which are eigenstates of the magnitude of the total 
angular-momentum operator, $J^2$, its $z$-component, $J_z$, 
the magnitude
of the orbital angular-momentum operator, $L^2$, and the
magnitude of the spin operator, $S^2$.
The partial wave amplitude can be written in terms of the original amplitude 
as
\beq
A^{\bar w' w}_{\ell \ell ' j}(q'_0,q';q_0,q;s) = {1 \over N(q',q) }
\int d \boldhp{q}
d \boldh{q} \, {\cal Y}^{\dagger}_{\ell  j m} (\boldhp{q})
\tilde{A}^{\bar w' w}
(q'_0,\boldp{q};q_0,\bold{q};s) {\cal Y}_{\ell ' j m} (\boldh{q}) \, .
\eeq

Applying the partial wave decomposition to the BS equation and making
use of the orthogonality of the generalized Legendre polynomials,
we obtain
\bea
T_{\ell j I}^{ \bar u u}(q'_0,q';q_0,q;s) & = & 
V_{\ell j I}^{ \bar u u}(q'_0,q';q_0,q;s) \nonumber \\
& &  \hspace{-3.5cm} + \, i  \sum _{w''=u,v}
\int_{-\infty}^{\infty}dq''_0 \int_0^{\infty}dq''
\, V_{\ell \ell' j I}^{ \bar u w''}(q'_0,q';q''_0,q'';s)
G^{\bar w'' w''}(q''_0,q'';s)T_{\ell' \ell j I}^{ \bar w'' u}
(q''_0,q'';q_0,q;s) \label{eq:pwbsea1} \, , \\
T_{\ell' \ell j I}^{ \bar v u}(q'_0,q';q_0,q;s) & = & 
V_{\ell' \ell j I}^{ \bar v u}(q'_0,q';q_0,q;s) \nonumber \\
& &  \hspace{-3.5cm} +\,  i  \sum _{w''=u,v}
\int_{-\infty}^{\infty}dq''_0 \int_0^{\infty}dq''
\, V_{\ell' \ell j I}^{ \bar v w''}(q'_0,q';q''_0,q'';s)
G^{\bar w'' w''}(q''_0,q'';s)T_{\ell' \ell j I}^{ \bar w'' u}
(q''_0,q'';q_0,q;s) \label{eq:pwbsea2} \, .
\eea
Note that the amplitude $A^{\bar u u}_{\ell \ell ' j}$ 
is diagonal in $\ell$, i.e.
$\ell = \ell '$, due to parity conservation, however amplitudes such
as $A^{\bar u v}_{\ell \ell ' j}$, which involve transitions 
between positive and negative energy nucleon states, 
are not diagonal in $\ell$.
Each partial wave amplitude $T^{\bar u u}$ is labelled by the orbital angular 
momentum $\ell$, the total angular momentum $j$, where
$j = \ell \pm \half$, and the total isospin $I$, with $I=\half$ or
$I=\thalf$.

\subsection{Analytic structure} \label{sec.6}

At this stage it is necessary to examine the analytic structure
of the partial wave BS equations. We will carry out a Wick
rotation~\cite{wick}, and analytically continue the $\pi N$ amplitude
in the $q'_0$ and $q''_0$ variables from the
real axis to the imaginary axis. Before doing this, we must examine
the singularity structure of the kernels of Eqs.~(\ref{eq:pwbsea1}) and
(\ref{eq:pwbsea2}) in the $q''_0$ plane, to make sure that 
there are no poles or cuts that could interfere
with the Wick rotation.
The residues of any poles present in the first
and third quadrants need to be picked up, since we rotate the
$q''_0$ integration contour from the real axis to the imaginary
axis in an anti-clockwise direction. The
presence of form factors in the potential ensures that 
the kernel is well behaved asymptotically, and as a result there is
no contribution from the contour at infinity. There are three
sources of analytic structure that we need to examine: (i) the
$\pi N$ intermediate state, 
(ii) the potential, in which there are poles
from the both the exchange particle propagators and the form
factors, and (iii) the $\pi N$ $T$-matrix itself.

The poles of the $\pi N$ two-body propagator
 $G^{\bar u u}(q_0'',q'';s)$ 
in the complex $q''_0$ plane are located at
\bea
q''_0 = \omega _N^+(q'')& \equiv & -\sqrt{s}/2 + \sqrt{q''^2+m_N^2} 
- i \epsilon   \, , \\
q''_0 = \omega _{\pi}^{\pm}(q'') & \equiv & \sqrt{s}/2 \mp 
\left( \sqrt{q''^2+\mpi^2} - i \epsilon \right)  \, ,
\eea
corresponding to the positive energy nucleon pole, and the positive and
negative energy pion poles respectively.
$G^{\bar v v}(q''_0,q'';s)$ has poles at $q''_0 = \omega _{\pi}^{\pm}$, and at
\beq
q''_0 = \omega _N^-(q'') \equiv  -\sqrt{s}/2 - \sqrt{q''^2+m_N^2} +
 i \epsilon  \, ,
\eeq
which corresponds to the negative energy nucleon pole.
If $\sqrt{s} > 2 \mpi$ it is possible for $\omega_{\pi}^+$
to be in the first quadrant for $0<q''<q''_{\mbox{\scriptsize max}}$,
where $q''_{\mbox{\scriptsize max}}=\sqrt{s/4-\mpi^2}$.
The positive energy nucleon pole $\omega_N^+$ can move into the third quadrant
for $\sqrt{s} > 2 m_N$. Therefore, if we stay below CM energies
of $2 m_N$, we only need to pick up the residues from the 
positive energy pion propagator pole.
The residues of $G^{\bar u u}(q''_0,q'';s)$
and $G^{\bar v v}(q''_0,q'';s)$ at $q''_0 = \omega _{\pi}^+(q'')$ are 
\bea
G^{\bar u u}_{\mbox{\scriptsize res}}(q'';s) &=&
-{m_N \over 2 E_{q''} \omega _{q''}} {1 \over 
\left( \sqrt{s} - E_{q''} - \omega _{q''} + i \epsilon \right)} \, , \\
G^{\bar v v}_{\mbox{\scriptsize res}}(q'';s)&=&
-{m_N \over 2 E_{q''} \omega _{q''}}
{1 \over \left( \sqrt{s} + E_{q''} - \omega _{q''} \right)} \, .
\eea

We now consider the singularities of all the diagrams present in the potential.
Firstly, note that the partial wave potentials have the form
\bea
V^{\bar u u}_{\ell j I} (q'_0,q';q''_0,q'';s) &= &
{ \pi \over N(q',q'')} \int_{-1}^{1}dx
\left( P_{\ell}(x)f^{\bar u u}_1(s,t,u) + {q'q'' \over \epsilon _{q'} \epsilon _{q''}} \, 
P_{\ell \pm 1} (x) f^{\bar u u}_2(s,t,u) \right), \label{eq:gfop1}  \\
V^{\bar u v}_{\ell \ell ' j I} (q'_0,q';q''_0,q'';s) &= &
{ \pi \over N(q',q'')} \int_{-1}^{1}dx
\left( {q'' \over \epsilon _{q''}} P_{\ell}(x) f^{\bar u v}_1(s,t,u) +
{q' \over \epsilon _{q'}} P_{\ell '}(x) f^{\bar u v}_2(s,t,u) \right), \label{eq:gfop2} 
\eea
where $x= \boldhp{q} \cdot \boldhpp{q}$ and
$\epsilon _q = E_q + m_N$. The forms of the partial wave potentials 
corresponding to $V^{\bar v v}$ and $V^{\bar v u}$ are 
very similar to Eqs.~(\ref{eq:gfop1}) and (\ref{eq:gfop2}) respectively.
For the $s$-channel baryon pole diagrams 
we have $f^{\bar w w}_i \propto 1/(s-m_B^2)$
with $i=1$, $2$. Therefore,
the only analytic structure in the $q''_0$ complex plane produced
by the $s$-channel pole diagrams is due to the form factors on the external
pion and nucleon legs, which we will look at shortly. However, 
for the $u$- and $t$-channel diagrams we have
$f^{\bar w w}_i \propto 1/(z-m^2)$,
where $z=u$ or $t$, and so the functions $f^{\bar w w}_i$ in this case depend on $q''_0$.
After carrying out the $x$ integration in 
Eqs.~(\ref{eq:gfop1}) and (\ref{eq:gfop2}),
the partial wave potentials corresponding to the
$u$- and $t$-channel pole diagrams will involve terms such as
\beq
\log \left( { (q'_0 + \eta q''_0)^2 - (q' + \eta q'')^2 
- m^2 + i \epsilon \over
(q'_0 + \eta q''_0)^2 - (q' - \eta q'')^2 
- m^2 + i \epsilon} \right) \, ,
\eeq
where $m$ is the mass of the particle, and $\eta = 1$ ($-1$) for the 
$u$-channel ($t$-channel) poles. Terms such as 
these generate logarithmic branch cuts in the $q''_0$ plane.  For the
$u$-channel pole diagrams, the branch points are at
\beq
q''_0 =- q_0' 
\pm \left( \sqrt{(q' \pm q'')^2 + m_B^2} - i \epsilon \right) \label{eq:bc1} \, ,
\eeq
where $B=N$ or $\Delta$.
The analytic structure for the
$t$-channel exchange diagrams is very similar, namely there are
branch points at
\beq
q_0'' = q_0' \pm 
\left( \sqrt{(q' \pm q'')^2 + m_A^2} - i \epsilon \right) \label{eq:bc2} \, ,
\eeq
where $A=\rho$ or $\sigma$. 

Notice that in Eqs.~(\ref{eq:bc1}) and (\ref{eq:bc2}) the positions of the branch points in the $q''_0$ plane 
depend on the external variable $q'_0$, unlike the poles of the
$\pi N$ intermediate state. Singularities
that depend on $q'_0$ are removed by the Wick rotation, since as well as rotating
the $q''_0$ integration contour from the real to the imaginary axis, we analytically
continue the $\pi N$ amplitude to the imaginary axis in $q'_0$. 
The branch cuts move away from the integration contour as we rotate from the
real to the imaginary axis, and in fact the branch points always stay
a distance $m$ away from the integration contour, where
$m$ is the mass of the exchange particle.

Next we consider the singularities of the form factors. The form
factors corresponding to the external pion and nucleon legs have
$n$-th order poles at
\bea
q''_0 &=& -\sqrt{s}/2
\pm \left( \sqrt{q''^2 + \Lambda_N^2} - i \epsilon \right) \, , \\
q''_0 &=& \sqrt{s}/2
\pm \left( \sqrt{q''^2 + \Lambda_{\pi}^2} - i \epsilon \right) \, ,
\eea
when the functional form given in Eq.~(\ref{eq:formI}) is used. The order
of the poles depends on the choice for the form factor powers, i.e.
$n_{\pi}$ and $n_N$.
To guarantee that these form factor poles do not interfere with the Wick 
rotation, we require that they do not move into the first or 
third quadrants of the $q''_0$ plane.
This means that we must have
$\sqrt{s} < 2 \Lambda_{\pi}$ and $\sqrt{s} < 2 \Lambda_N$.
Note that if form factors of the type given in Eq.~(\ref{eq:formII}) were
being used, there would be poles from each form factor
in both the first and third quadrants.

As a 
next step we look at the singularities in the form factors 
corresponding to the exchange particles. For the $u$-channel exchange diagrams
there are branch points at
\beq
q''_0 =- q_0' 
\pm \left( \sqrt{(q' \pm q'')^2 + \Lambda_B^2} - i \epsilon \right) \, , \label{eq:bc3}
\eeq
where again $B=N$ or $\Delta$. Finally for the
$t$-channel diagrams there are branch points at
\beq
q_0'' = q_0' \pm
\left( \sqrt{(q' \pm q'')^2 + \Lambda_A^2}- i \epsilon \right) \, , \label{eq:bc4}
\eeq
where $A=\rho$ or $\sigma$. All of these branch points are removed by the Wick
rotation, due to the $q'_0$ dependence.

The remaining source of singularities which must be considered
is the $T$-matrix, i.e. the solution of the BS equation.
This can be done by looking at what happens as the 
potential is iterated in the BS equation. Singularities in the
$q'_0$ plane of $T(q',q;P)$ are generated by pairs of poles pinching the 
$q''_0$ integration contour. This analytic structure therefore
also occurs in the $q''_0$ plane of $T(q'',q;P)$ which appears in the
integrand of the BS equation. As the BS equation is iterated a
hierarchy of branch cuts are generated. Higher order branch cuts 
arise from the pinching between lower order branch cuts and the 
singularities of the $\pi N$ intermediate state and potential. The positions
of most of these cuts depends on $q'_0$, and so do not cross the
integration contour when both the $q'_0$ and $q''_0$ axes are rotated.
However, some of the higher order branch cuts do not depend on
$q'_0$ and so are not removed by a Wick rotation.
If any pair of these cuts in the $q''_0$ plane protrude 
into both the first and third quadrants simultaneously,
a simple Wick rotation becomes no longer possible.

Also, singularities pinching the $q''_0$
integration contour can produce cuts in the $\sqrt{s}$ plane,
i.e.  thresholds. If these thresholds are generated by singularities
other than those from the form factors,  they correspond to physical
processes. The lowest-energy physical thresholds are at:
\bea
\sqrt{s} & = & m_N + m_{\pi} \nonumber \, ,\\
\sqrt{s} & = & m_N + 2 m_{\pi} \nonumber \, ,\\
\sqrt{s} & = & m_{\Delta} + 2 m_{\pi} \nonumber \, ,\\
\sqrt{s} & = & m_N + m_{\pi} + m_{\sigma} \nonumber \, .
\eea
However, if a cut in the $\sqrt{s}$-plane is generated 
by the pinching of a form factor pole and another singularity, the 
threshold is unphysical, since the cutoff mass does not correspond to
the mass of a physical particle. Some examples of the unphysical 
thresholds generated by the BS equation include:
\bea
\sqrt{s} & = & \Lambda _N + m_{\pi} \nonumber \, ,\\
\sqrt{s} & = & m_N + \Lambda _{\pi} \nonumber \, ,\\
\sqrt{s} & = & \Lambda _N + \Lambda _{\pi} \nonumber \, ,\\
\sqrt{s} & = & m_N + m_{\pi} + \Lambda_{\sigma} \nonumber \, .
\eea
We therefore have to make sure that 
the cutoff masses are chosen to be
large enough so that the unphysical thresholds occur 
above the highest CM energy for which we will solve the BS equation.

In summary,
the BS equation can be solved for $\pi N$ scattering with our choice of form factors
using a Wick rotation,
provided that the cutoff masses are not too small. There are three
conditions on the minimum values of the cutoff masses: 
(i) there are no form factor poles in the first
or third quadrants of the $q''_0$ plane, 
(ii) any cuts 
produced by the pinching between form factor poles and other
singularities are also not in the first or third quadrants, and finally
(iii) all unphysical thresholds, which are generated by the form factors, 
are far away from the energy region in which we are interested.
The minimum values for the cutoff masses that can be used in the BS
equation are given in Table~\ref{minall}. Here we have assumed that 
the two-pion production threshold is the maximum CM energy for
which we will solve the BS equation.

If form factors of the type given in Eq.~(\ref{eq:formII})
are used, it is not possible to prevent the
form factor singularities from interfering with the Wick rotation by making
any appropriate choices for the cutoff masses. There will always be poles
from the form factors in each quadrant of the $q''_0$ plane.

\subsection{Wick rotation}

Having looked at the analytic structure of the kernel of the partial wave 
BS equation, we are now in a position to perform a Wick rotation on
Eqs.~(\ref{eq:pwbsea1}) and (\ref{eq:pwbsea2}), by making the substitutions
\beq
q'_0 \rightarrow iq'_0 \, , \hspace*{1.2cm}
q''_0 \rightarrow iq''_0 \, ,
\eeq
and picking up the residues from any poles in the first or third
quadrants of the complex $q''_0$  plane. After Wick rotation,
the partial wave BS equation becomes a system of four coupled
integral equations. The first two equations for the half-off-shell
$T$-matrix are
\bea
T_{\ell j I}^{ \bar  u u}(iq'_0,q'; \bar q_0,  \bar q;s) & = & 
V_{\ell j I}^{ \bar  u u}(iq'_0,q'; \bar q_0, \bar q;s) \nonumber \\
& &  \hspace{-3.5cm} -  \sum _{w''=u,v}
\int_{-\infty}^{\infty}dq''_0 \int_0^{\infty}dq''
\, V_{\ell \ell' j I}^{ \bar  u w''}(iq'_0,q';iq''_0,q'';s)
G^{ \bar w'' w''}(iq''_0,q'';s)T_{\ell' \ell j I}^{ \bar  w'' u}
(iq''_0,q''; \bar q_0, \bar q;s) \nonumber \\
& &  \hspace{-3.5cm} -  \sum _{w''=u,v}
\int_0^{q''_{\mbox{\scriptsize max}}}dq''
\, V_{\ell \ell' j I}^{  \bar u w''}
(iq'_0,q';\omega_{\pi}^+(q''),q'';s)
G^{\bar  w'' w''}_{\mbox{\scriptsize res}}(q'';s)
T_{\ell' \ell j I}^{  \bar w'' u}
(\omega_{\pi}^+(q''),q''; \bar q_0, \bar  q;s) \, , \label{eq:pwbsew1} \\
T_{\ell' \ell j I}^{ \bar  v u}(iq'_0,q'; \bar q_0, \bar q;s) & = & 
V_{\ell' \ell j I}^{ \bar  v u}(iq'_0,q'; \bar q_0, \bar q;s) \nonumber \\
& &  \hspace{-3.5cm} -  \sum _{w''=u,v}
\int_{-\infty}^{\infty}dq''_0 \int_0^{\infty}dq''
\, V_{\ell' \ell j I}^{  \bar v w''}(iq'_0,q';iq''_0,q'';s)
G^{ \bar w'' w''}(iq''_0,q'';s)T_{\ell' \ell j I}^{ \bar  w'' u}
(iq''_0,q''; \bar q_0, \bar q;s) \nonumber \\
& &  \hspace{-3.5cm} -  \sum _{w''=u,v}
\int_0^{q''_{\mbox{\scriptsize max}}}dq''
\, V_{\ell' \ell j I}^{ \bar  v w''}
(iq'_0,q';\omega_{\pi}^+(q''),q'';s)
G^{ \bar w'' w''}_{\mbox{\scriptsize res}}(q'';s)
T_{\ell' \ell j I}^{ \bar  w'' u}
(\omega_{\pi}^+(q''),q''; \bar q_0, \bar q;s) \, . \label{eq:pwbsew2}
\eea
In the above we have put the pion and nucleon in the initial state
on-mass-shell. The on-shell relative momenta are denoted by 
$\bar q_0$ and $\bar q$, and are given by
\beq
\bar q _0 = {1 \over 2} \left( \sqrt{\bar q^2 + m_N^2} -
\sqrt{\bar q ^2 + \mpi^2} \right) \, , \label{eq:onshell}
\eeq
and
\beq
\bar q = \sqrt{[s-(m_N+m_{\pi})^2][s-(m_N-m_{\pi})^2] \over 4s} \, .
\eeq
There are two additional equations
(usually referred to as the ``auxiliary equations'') which are necessary
in order to have a closed system of equations to solve: they are
Eqs.~(\ref{eq:pwbsew1}) and (\ref{eq:pwbsew2})
with  $iq'_0$ replaced with $\omega_{\pi}^+(q')$.

Finally, it is necessary to look at each term present in the potentials to check
whether there are any remaining singularities after Wick rotation. 
For energies above the pion production threshold, the $u$-channel nucleon
pole present in the potential in the one-dimensional parts of the auxiliary equations
develops an imaginary part. This is due to a logarithmic singularity moving into
the integration region, and must be handled carefully to ensure numerically
stable results. We do this by carrying out a subtraction similar to Ref.~\cite{grossnn}.
There are no additional singularities
caused by the form factors below the two-pion production threshold,
provided the cutoff masses are chosen to be
larger than the values given in Table~\ref{minall}.

Above $\sqrt{s} = 2( m_{\sigma} + m_{\pi})$ a cut in the $q''_0$ plane,
generated by the pinching of the integration contour between the 
positive-energy pion pole and the $\sigma$ meson propagator, moves into the
1st quadrant.  Therefore, above this value of the CM energy, it 
would become necessary to take this additional singularity into account 
when carrying out the Wick rotation. 
Here we consider CM energies below $\sqrt{s} = 2( m_{\sigma} + m_{\pi})$.

\subsection{Calculation of the phase shifts}

In order to determine the $\pi N$ phase shifts, the on-shell amplitude 
$T^{\bar u u}_{\ell j I}(\bar q_0, \bar q; \bar q_0, \bar q;s)$  needs to
be calculated. This is done by analytic continuation of the half-off-shell
amplitude to the on-shell point. In practice, the on-shell $T$-matrix is
obtained using Eq.~(\ref{eq:pwbsew1}), with both the incoming
and outgoing particles on-mass-shell.

The $\pi N$ two-body propagator 
$G^{\bar u u}_{\mbox{\scriptsize res}}(q'';s)$ has a pole 
when the pion and nucleon are propagating on-shell, 
which occurs for
$q''=\bar q$. This pole is related to  the two-body
unitarity cut. 
For energies above the $\pi N$ threshold we need to take care of this pole
so as to obtain equations that can be solved
numerically. We achieve this by writing 
 $G^{\bar u u}_{\mbox{\scriptsize res}}(q'';s)$ in terms of a principal-value
part and an imaginary on-shell contribution, i.e.
\beq
G^{\bar u u}_{\mbox{\scriptsize res}}(q'';s)=
-{m_N \over 2 E_{q''} \omega _{q''}}
{{\cal P} \over
\left( \sqrt{s} - E_{q''} - \omega _{q''} \right)}
+ {i \pi m_N \over 2 \bar q \sqrt{s}} \, \delta(q'' - \bar q) \, , \label{eq:greenpv}
\eeq
where ${\cal P}$ denotes that the principal-value prescription should be
used when the $q''$ integration is performed.

The $\pi N$ phase shifts $ \delta _{\ell j I}$ and the inelasticities
$\eta _{\ell j I}$ are obtained from the on-shell partial wave $T$-matrix
using
\beq
T^{\bar u u}_{\ell j I}(\bar q_0, \bar q; \bar q_0, \bar q;s)
= - {\bar q \sqrt{s} \over m_N \pi^2} \,
\left( {\eta _{\ell j I} e^{2 i \delta _{\ell j I}} - 1 \over
2 i \bar q} \right) \, .
\eeq
The behaviour at threshold is more conveniently described in terms
of scattering lengths and volumes, which are defined by the 
effective range expansion:
\beq
\bar q^{2 \ell + 1} \cot \delta _{\ell j I} = {1 \over a_{\ell j I}} 
+ {1 \over 2} \, r_{\ell j I} \, \bar q^2 + \ldots \, .
\eeq
Here $a_{\ell j I}$ is the scattering length and $r_{\ell j I}$ is 
the effective range.

\section{NUMERICAL RESULTS}\label{sec.7}

\subsection{Fits to the empirical $\pi N$ data}

We begin this section by listing the free parameters in our model. The five cutoff
masses ($\Lambda _{N}$, $\Lambda _{\Delta}$, $\Lambda _{\pi}$, 
$\Lambda _{\rho}$ and $\Lambda _{\sigma}$) are free, but are constrained
to be larger than the minimum values given in Table~\ref{minall}. Due
to the uncertainty in the values of the coupling constants other than
$g_{\pi NN}$, we permit $f_{\pi N \Delta}$, $x_{\Delta}$, $g_{\rho}$,
$\kappa _{\rho}$ and $g_{\sigma \pi \pi} g_{\sigma N N}$ to vary freely.
The mass of the $\sigma$ meson, the bare $\Delta$ mass $m_{\Delta}^{(0)}$ 
and the bare $\pi N \Delta$ coupling constant $f_{\pi N \Delta}^{(0)}$ are also 
allowed to vary freely, although $m_{\Delta}^{(0)}$ and $f_{\pi N \Delta}^{(0)}$
are essentially fixed by the $P_{33}$ phase shifts. The bare nucleon mass
$m_N^{(0)}$ and the bare $\pi N N$ coupling constant $g_{\pi N N}^{(0)}$ are not
free parameters, but are determined by the renormalization procedure outlined in 
Section~\ref{sec.4}. We use the value $n_a=1$ for all the form factor powers,
however it turns out that this choice is not crucial to the quality of the fit.

\begin{figure}
\begin{center}
\epsfig{figure=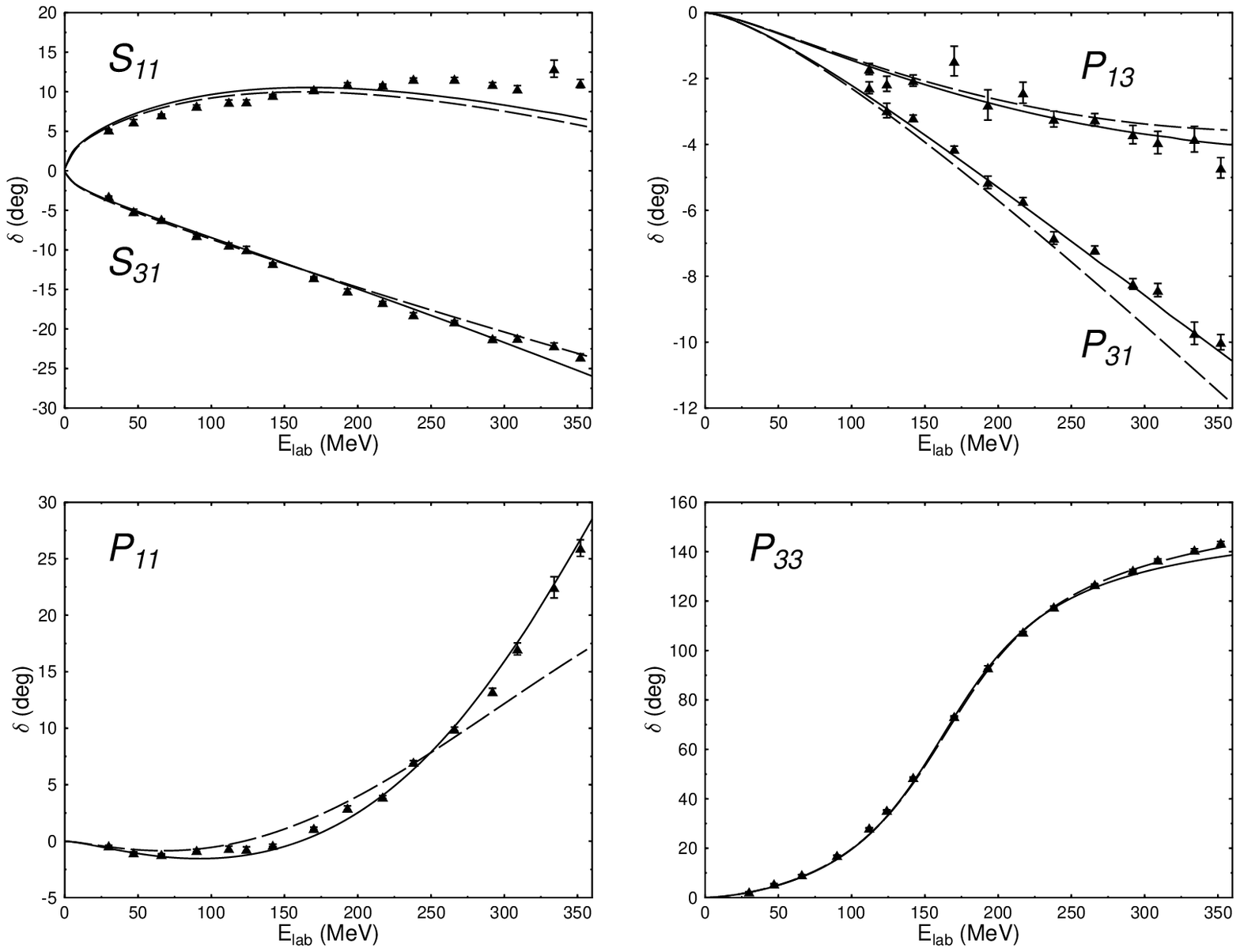,width=14.5cm,scale=0.5,angle=-90}
\end{center}
\vspace*{-4.5cm}
\caption{The phase shifts obtained from the BS equation
shown versus the pion laboratory energy, using the Rarita-Schwinger
$\Delta$ propagator (\solid) and Pascalutsa $\Delta$ propagator
(\dash). Data points from the VPI SM95 partial wave analysis are also shown.}
\label{fig:p1}
\end{figure}

The free parameters are determined in $\chi ^2$ fits to the $s$- and $p$-wave 
single-energy phase shifts up to 360 MeV pion laboratory energy, as well as
the scattering lengths and volumes, from the VPI SM95 partial wave 
analysis~\cite{sm95}. We carry out one fit using the Rarita-Schwinger 
$\Delta$ propagator, and another using the Pascalutsa
propagator. The coupling constants and particle masses for both fits are 
listed in 
Table~\ref{parms}. The resulting scattering lengths and volumes are given 
in Table~\ref{scatt1}, and the phase shifts are shown in Fig.~\ref{fig:p1}.
We see that the BS equation gives a good description of the $\pi N$
phase shifts. Notice that the results for the phase shifts in the $P_{11}$ 
partial wave are better when the Rarita-Schwinger $\Delta$ propagator is used.

As can be seen in Table~\ref{parms}, the values of 
$g_{\sigma \pi \pi} g_{\sigma N N}$ determined from both fits have 
negative signs,
which means that the $\sigma$ contribution is repulsive in the $s$-waves
and attractive in the $p$-waves, as was also found in 
Refs.~\cite{schutz94,tjon98}. 
Note that we have used a low value for the
$\pi N N$ coupling constant, i.e. $g_{\pi NN}^2 / 4 \pi = 13.5$. We
have repeated the fits using $g_{\pi NN}^2 / 4 \pi = 14.3$
and found that the results are very similar to those shown in Fig.~\ref{fig:p1}.
Also, there are no significant differences between the coupling constants 
obtained from the fits using
$g_{\pi NN}^2 / 4 \pi = 13.5$ and $g_{\pi NN}^2 / 4 \pi = 14.3$.

In Table~\ref{comps} we compare our coupling constants to those extracted
from $\pi N$ models based on the $K$-matrix approximation and 3-D reductions
of the BS equation. Our coupling constants are in general consistent
with those obtained from other models of $\pi N$ scattering.
We note that $f_{\pi N \Delta}^2/ 4 \pi$ is in the range 0.35 to 0.43 for all 
equations, except for the BS equation when the Pascalutsa $\Delta$ propagator
is used, in which case the $\pi N \Delta$ coupling constant is about twice
as large as the commonly accepted value.
Our value of $x_{\Delta}$ is similar to those found in 
other models, and they are all in the range $-0.41 < x_{\Delta} < -0.11$.
All the models listed give similar values for $g_{\rho}^2$, and
the values of $g_{\rho}^2$ all lie in the range 
$2.5 < g^2_{\rho} / 4 \pi < 3.36$. Our values are consistent with
the KSRF relation~\cite{ksrf} and the value found from the width of the pionic decay 
of the $\rho$ meson. 
There is a large range of values for $\kappa _{\rho}$, which vary between 
1.44 and 6.6. Our value of $\kappa _{\rho} = 2.66$ using the RS propagator 
is  smaller than the VMD result of $\kappa _{\rho} = 3.7$, and we get a 
value slightly larger than the VMD result ($\kappa _{\rho} = 4.11$) when we use the 
Pascalutsa $\Delta$ propagator. This suggests that within the uncertainty
from the $\Delta$ propagator, we are consistent with vector meson
dominance. 

While we are to get a better fit to the $\pi N$ phase shifts using the
Rarita-Schwinger $\Delta$ propagator than when using the Pascalutsa propagator,
this does not necessarily suggest that the Rarita-Schwinger propagator 
is the correct spin-3/2 propagator. Other processes not included in the present model
can give attractive contributions to the $P_{11}$ partial wave, such as
the coupling to the $\pi \Delta$ channel or the inclusion of the $N^*(1440)$
resonance into the potential. 
More important is the observation that the different choices for the $\Delta$
propagator give rise to differences in the coupling constants. This 
highlights the importance of having a better understanding of how to construct
propagators for higher-spin particles.

Having compared results using two different choices for the $\Delta$ propagator, 
hereafter we restrict ourselves to the Rarita-Schwinger propagator.

\subsection{Contributions to the phase shifts}

\begin{figure}
\begin{center}
\epsfig{figure=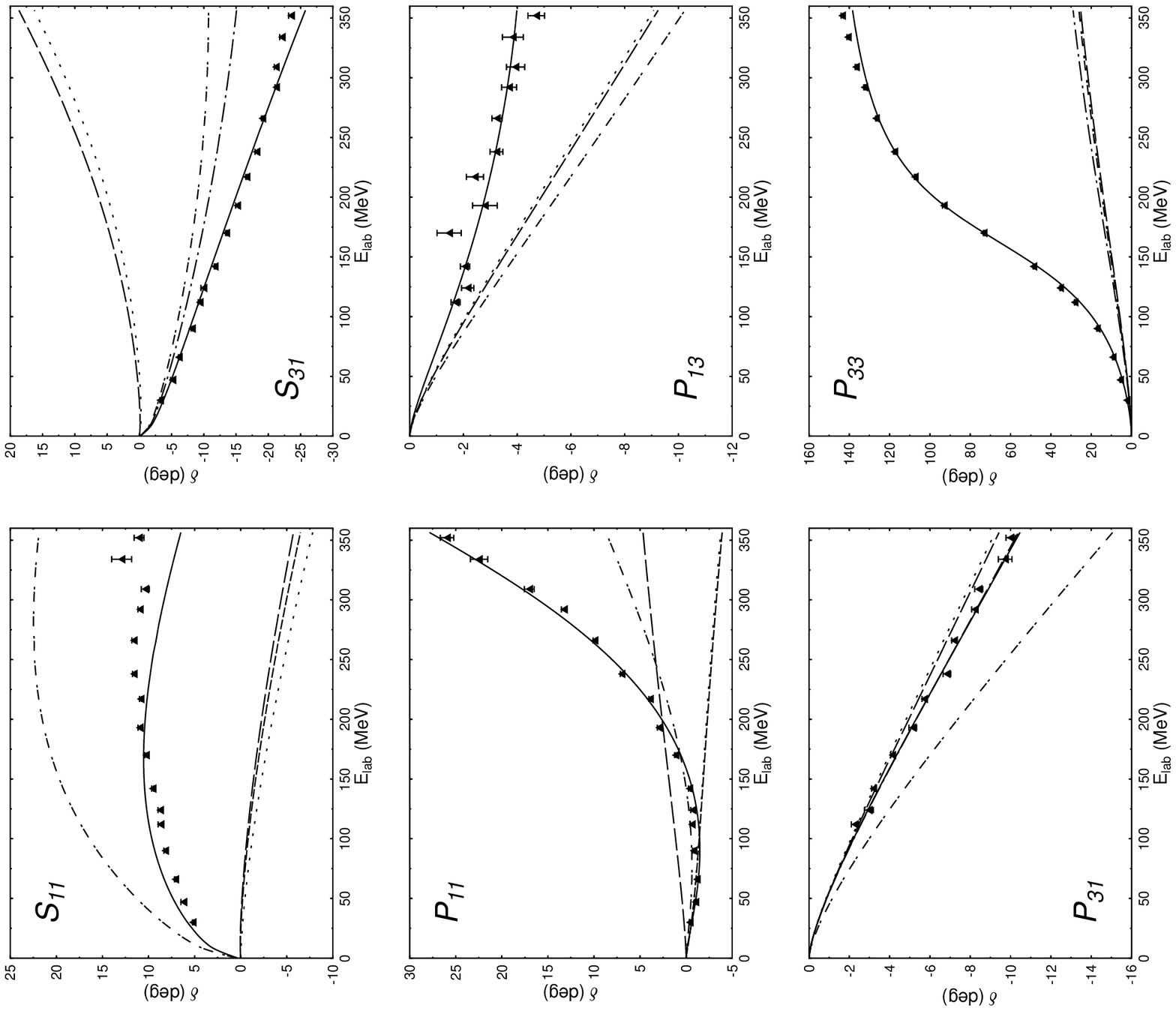,width=14.5cm,scale=0.5,angle=-90}
\end{center}
\vspace*{0.0cm}
\caption{Contributions to the $\pi N$ phase shifts as each diagram
is added to the potential, in the following order:
$u$-channel $N$ pole (\dash),
$s$-channel $N$ pole (\dashes), 
$t$-channel $\sigma$ exchange (\dotss),
$t$-channel $\rho$ exchange (\dashdot), 
$u$-channel $\Delta$ pole (\dashdash), and 
$s$-channel $\Delta$ pole (\solid). \label{fig:contr}}
\end{figure}

In Fig.~\ref{fig:contr} we show how the total phase shifts 
are built up from the contributions of the individual 
Feynman diagrams in the potential. The
$u$-channel nucleon pole is strongest in the $S_{31}$ and $P_{33}$ partial
waves, but also gives important contributions to $P_{13}$ and $P_{31}$. The
$s$-channel nucleon pole generates the repulsion in the $P_{11}$ phase
shifts, and gives a very small contribution to $S_{11}$. The $u$-channel
$\Delta$ diagram plays a very important role in all partial waves except
$S_{31}$ and $P_{33}$, as it gives a large repulsive contribution to $S_{11}$,
and gives strong attractive contributions to $P_{11}$, $P_{13}$ and
$P_{31}$. The $s$-channel $\Delta$ pole diagram dominates the $P_{33}$ phase shifts,
but also gives an important contribution to the $S_{31}$ partial wave,
and a tiny contribution to $P_{31}$. The contributions from the $s$-channel
$\Delta$ pole to $S_{31}$ and $P_{31}$ result from the spin-1/2 components
of the Rarita-Schwinger propagator. The $\rho$ exchange contributions are
largest in the $s$-waves, but are also significant in the $p$-waves except
for $P_{33}$. Likewise, $\sigma$ exchange is stronger in  the $s$-waves than
the $p$-waves, although the $\sigma$ exchange contributions are quite small 
in all partial waves.

The $S_{11}$ phase shifts are dominated by $\rho$ exchange and the
$u$-channel $\Delta$ pole. By itself, the attraction generated by
$\rho$ exchange is far too strong, but is partially cancelled by the
repulsive $u$-channel $\Delta$ pole.
Except for the $u$-channel nucleon pole, all of the diagrams contributing
to the $S_{31}$ phase shifts are repulsive. The largest contributions to
this repulsion come from $\rho$ exchange and the spin-1/2 components of the
$s$-channel $\Delta$ pole. 

The $s$-channel nucleon pole causes the $P_{11}$ phase shifts to be 
negative at low energies.
The attraction that causes the phase shifts
to change sign is dominated by the $u$-channel $\Delta$ pole and the
$\rho$ exchange diagram.

The $u$-channel nucleon pole is strong and repulsive in the $P_{13}$ and
$P_{31}$ partial waves. In fact, by itself, the $u$-channel nucleon pole
almost gives the correct $P_{31}$ phase shifts. However, the 
$\rho$ exchange diagram gives a large repulsive contribution to $P_{31}$,
causing the phase shifts to deviate strongly from the phase shift analysis.
The attraction provided by the $u$-channel $\Delta$ pole to $P_{31}$
almost cancels the $\rho$ contribution completely. Similarly, the $P_{13}$
phase shifts are far too repulsive without the strong attraction produced
by the $u$-channel $\Delta$ pole.

The $P_{33}$ phase shifts are of course dominated by the $s$-channel
$\Delta$ pole, with the background contribution primarily coming from
the $u$-channel nucleon pole.

\subsection{Dressing of the $\pi N N$ vertex}

As can be seen in Table~\ref{parms}, the cutoff masses we obtain turn
out to be quite large. This results in the dressing being very
significant, as is evident from the large size of the bare $N$ and
$\Delta$ masses. 
In view of the significance of the dressing, it is interesting to
examine the effect of  dressing on the $\pi NN$ form factor. When
both nucleons in the bare $\pi N N$ vertex are placed on-mass-shell, the
bare $\pi N N $ vertex only involves the pion form factor, and is given by
\beq
\Gamma ^{(0) \bar u u} _{\pi N N}(q_0,\bold{q};s) = f _{\pi} (q_{\pi}^2)
{\cal V} ^{\bar u u} _{\pi N N}(q_0, \bold{q};s) \, ,
\eeq
where $s=m_N^2$, and
 $q^2_{\pi}$ is the 4-momentum squared of the pion. The bare
 pion form factor is
\beq 
f_{\pi }(q^2_{\pi})=
\left( { \Lambda ^2_{\pi} - m^2_{\pi}
\over \Lambda ^2_{\pi} - q^2_{\pi}} \right) ^{n_{\pi}} \, , \label{eq:noms}
\eeq
and we  take $n_{\pi} = 1$.
The renormalized pion form factor can then be introduced as
\beq
f^R_{\pi }(q^2_{\pi})= 
Z_1 \, { \Gamma ^{ \bar u u} _{\pi N N}(q_0,q;s) \over 
{\cal V} ^{\bar u u} _{\pi N N}(q_0,q;s)} \Biggl| _{P_{11}} \label{eq:rpff} \, ,
\eeq
which has the property that $f^R_{\pi }(m^2_{\pi})=1$, and
where $q_0$ and $q$ are related to the
pion 4-momentum squared by
\bea
q_0 & = & {1 \over 2 m_N} \left( m_N^2 - q_{\pi}^2 \right) \, , \\
q & = & {1 \over 2} \sqrt{q_{\pi}^2 
\left( {q_{\pi}^2 \over m_N^2} - 4 \right) } \, .
\eea
In Eq.~(\ref{eq:rpff}) we have taken the $P_{11}$ partial wave of 
the quantities $\Gamma ^{\bar u u} _{\pi N N}(q_0,\bold{q};s)$
and ${\cal V} ^{\bar u u} _{\pi N N}(q_0, \bold{q};s)$.
By comparing the slope of the function 
$f^R_{\pi }(q^2_{\pi})$ at $q^2_{\pi}=0$ with
a monopole form factor with cutoff mass $\Lambda ^R_{\pi}$,
we can obtain a value for the renormalized pion cutoff mass. We 
find that $\Lambda _{\pi}^R=1.22$ GeV, which is 
 softer than the bare pion form factor
(recall that $\Lambda _{\pi}=1.77$ GeV). This is consistent with previous
calculations~\cite{pearce,liu95,scholten98}, which have found that dressed
form factors are softer than the corresponding bare form factors.

We can introduce the quantity $\Delta _{\pi}$ as a measure of the variation 
between the renormalized pion form factor at $q_{\pi}^2 = m_{\pi}^2$ and 
$q_{\pi}^2 = 0$, i.e.
\beq
\Delta _{\pi} = 1 - f^R_{\pi}(0) \, .
\eeq
We find that $\Delta _{\pi} = 1.3 \%$, which indicates that our
dressed $\pi N N$ form factor is a very slowly-varying function of the
pion mass. Our value of $\Delta _{\pi}$ is somewhat smaller than the
value of $3 \%$ obtained using other methods~\cite{soft}.

\subsection{Different choice of form factors}

With the choice of form factors we have used so far 
(hereafter referred to as model I), the effect of 
dressing is significant. We now 
consider the case where there is a form factor only on the pion 
(referred to as model II), which is arrived at from the parameterization of 
form factors used in model I by taking the limit
$\Lambda _h \rightarrow \infty$ for $h=N$, $\Delta$, $\rho$, 
and $\sigma$.
With this choice of form factor there is only one cutoff mass, rather
than five, and so the number of free parameters is reduced by four. All 
intermediate states contain the pion propagator, and therefore a cutoff function 
still appears in all loop diagrams to provide convergence.
In model II the pion form factor is used to vary the off-mass-shell
behaviour of the pion. This in principle could be constrained by
the soft-pion theorems.

In Table~\ref{pionpars} we show three sets of parameters obtained 
from fits to the on-shell $\pi N$ data, corresponding to the choices
$n_{\pi} = 2$, $n_{\pi} = 4$, and $n_{\pi} = 10$. The resulting
phase shifts are all of the same quality as the results
using the Rarita-Schwinger $\Delta$ propagator shown in Fig.~\ref{fig:p1}.
The coupling constants resulting from the three model II fits
are similar to the coupling constants obtained using model I, although
$\kappa _{\rho}$ is smaller in model II than in model I, and 
$g_{\sigma \pi \pi} g_{\sigma NN}$ has a positive sign, whereas in
model I it is negative.
As with model I, there are no significant changes to the quality of the fits
or values of the coupling constants obtained when the fits are repeated
using $g_{\pi N N}^2 / 4 \pi = 14.3$ as the physical $\pi N N$ 
coupling constant.

The main difference between the two models is that
in model II  the effect of dressing is not as significant as in model I. 
The bare baryon masses are much closer to the physical masses. 
The renormalized pion cutoff masses and values of $\Delta _{\pi}$
for model II are given in Table~\ref{pionren}. We see that the values of
$\Lambda _{\pi}^R$ are smaller than in model I, and are close to the value of 
$\Lambda _{\pi}^R \approx 0.8$ GeV advocated by some authors~\cite{thomas8}.
The values of $\Delta _{\pi}$ are consistent with 
previous calculations~\cite{soft} of the difference between 
$f_{\pi}^R(m_{\pi}^2)$ and $f_{\pi}^R(0)$.

\subsection{Above the $2 \pi$ production threshold}

\begin{figure}
\begin{center}
\epsfig{figure=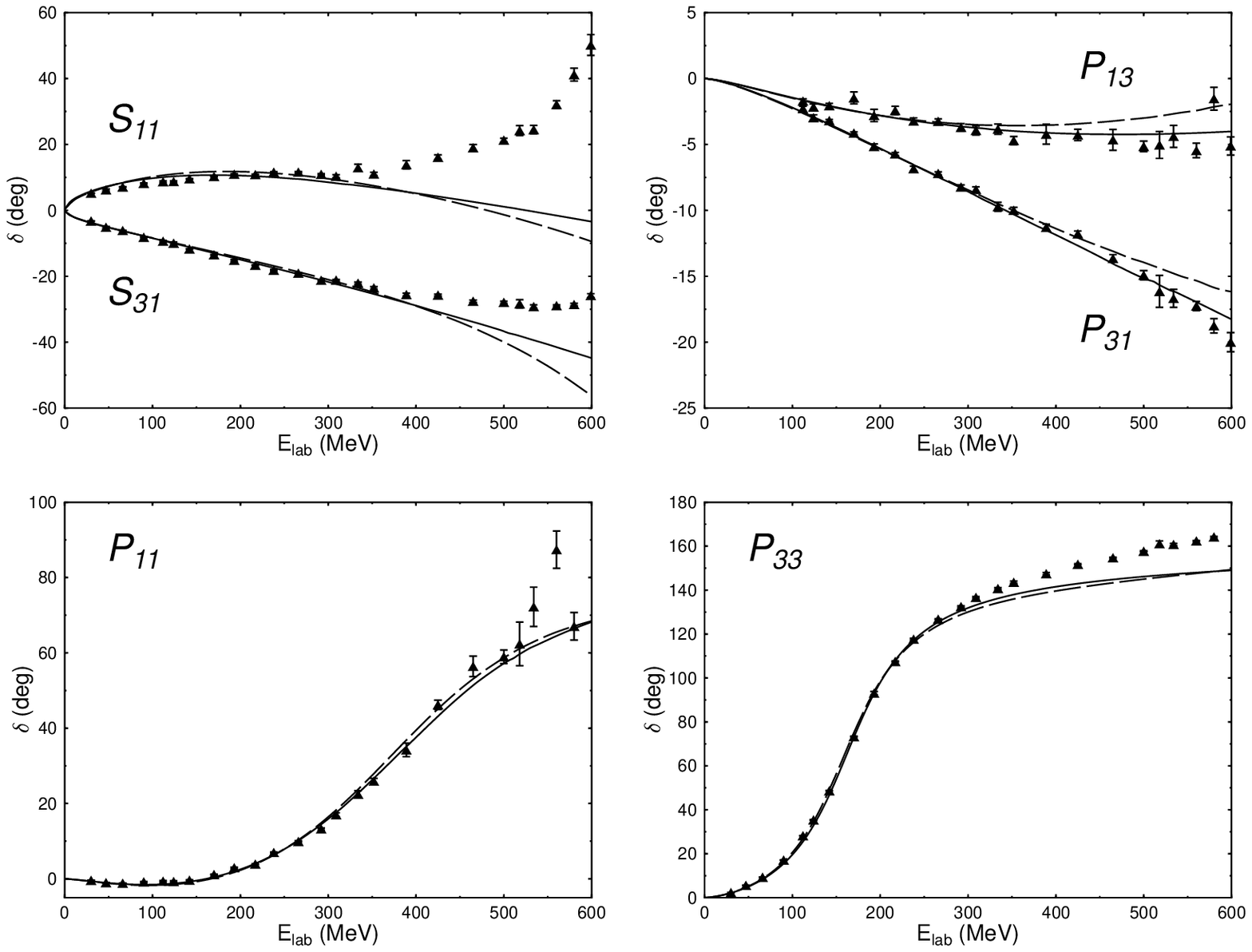,width=14.5cm,scale=0.5,angle=-90}
\end{center}
\vspace*{-4.5cm}
\caption{The phase shifts obtained from the BS equation
are shown versus the pion laboratory energy up to 600 MeV, for
models I (\solid) and II (\dash).}
\label{fig.ps600}
\end{figure}

In order to see what happens at energies above the $2 \pi$ production
threshold, we show the phase shifts up to 600 MeV pion laboratory energy
in Fig.~\ref{fig.ps600}. The results from model II were obtained using
$n_{\pi} = 4$. In the $P_{11}$ and $P_{33}$ partial waves both models
give almost identical results, but there are some differences between models
I and II in the higher energy region in the other partial waves.

Both the $P_{13}$ and $P_{31}$ phase shifts 
are quite good over the full range of energies, which is a reflection of the
fact that below 600 MeV there are no resonances in these partial waves and 
the inelasticity is negligible. 
We require a little more attraction at the higher energies in the
$S_{31}$ and $P_{33}$ partial waves, while a significant amount of 
additional attraction is required for $S_{11}$ above 300~MeV, and 
for $P_{11}$ above around 450~MeV.

It is not unexpected that there are some discrepancies in the higher energy 
region. Here the $S_{11}$ and $P_{11}$ partial waves exhibit resonance
behaviour not included in the present model. There are a number of
modifications that could be made to our model in order to improve the 
agreement with experiment for this larger energy range. Firstly, it may be 
necessary for three-body unitarity to be satisfied. Amongst other things, 
this will involve replacing the nucleon propagator in the $\pi N$ intermediate 
states with a dressed propagator. 
Extending the model to include the coupling to inelastic channels 
and the possible addition of explicit nucleon resonances into the potential
will be essential at energies above 360~MeV. To fully understand the $\pi N$
amplitude at these energies, the coupling to inelastic channels must first
be included, and if the fit to the phase shifts is still unsatisfactory, 
explicit bare $N^*$ baryon poles may need to be included in the potential.

The $S_{11}$ partial wave would be improved by the inclusion of the coupling
to the $\eta N$ channel, and also $S_{11}$ resonances such as the
$N^*(1535)$ and $N^*(1650)$ may need to be included in the potential. 
The coupling to the $\pi \Delta$ and $\sigma N$ channels 
and possibly the inclusion of
the $N^*(1440)$ resonance into the potential would be necessary to improve
the $P_{11}$ phase shifts above 450~MeV.

\section{CONCLUSION}\label{sec.8}

In this work we have presented a description of pion-nucleon 
scattering based on the four-dimensional Bethe-Salpeter equation. 
The kernel of the equation is based on a chiral Lagrangian that 
includes in addition to pions and nucleons, the $\Delta$(1232) and 
the $\rho$ and $\sigma$ mesons. The potential obtained from this
Lagrangian consists of $s$- and $u$-channel $N$ and $\Delta$ pole diagrams 
as well as $t$-channel $\rho$  and $\sigma$ exchanges. Convergence of
all integrals is guaranteed by the use of cutoff fuctions associated 
with each vertex.
Two different parameterizations of the cutoff functions were considered: 
in model I the cutoff function was taken to
be a product of form factors depending on the four-momentum squared of 
each particle present at the vertex. In model II the cutoff function was
taken to depend only on the pion four-momentum squared.
The parameters of the potential were adjusted to fit the empirical $s$- 
and $p$-wave phase shifts up to a pion laboratory kinetic energy of 360~MeV.
Both models give good fits to the $\pi N$ scattering data, and 
the resulting coupling constants are consistent with the commonly accepted 
values extracted from other observables. 

While most of our results were for the Rarita-Schwinger $\Delta$ 
propagator, we compared the results of fits performed using the Rarita-Schwinger 
and Pascalutsa propagators for the case of the model I form factors.
The differences in the coupling constants obtained suggests that a complete
understanding of the baryon resonances with higher-spin 
is not possible without having unique higher-spin propagators.

The good fits to the on-shell $\pi N$ data for energies below 360~MeV 
suggests that a model of the $\pi N$ interaction based on the Bethe-Salpeter
equation could form the basis for the analysis of pion 
photoproduction by the proper $U(1)$ gauging of the Lagrangian~\cite{aa95}, 
and the analysis of the baryon resonances near and above the threshold 
for pion production. In fact, by extending the calculations to pion 
energies up to 600~MeV, we observe that in partial waves that don't have 
large inelasticities and don't exhibit evidence of baryon states, 
both our models give good representations of the data.
For partial waves with large inelasticity there is also evidence for baryon 
resonances, and as a result we need to include first the coupling to 
inelastic channels, and then include bare baryon states into the potential
if necessary.

Although we have not included full three-body unitarity, which would
require the dressing of the nucleon propagators and $\pi N N$ vertices, 
we have included those contributions to three-body unitarity resulting 
from the fact that we have not carried out any three-dimensional reduction,
and as such our potentials depend on the relative energy. The inclusion
of the dressed nucleon in the $\pi N$ intermediate states will increase the 
number of coupled channels, and is under investigation. 

By calculating off-mass-shell $\pi N$ amplitudes 
using the Bethe-Salpeter equation we can examine the low energy
theorems, and study the questions associated with the analytic continuation 
of the physical $\pi N$ data to the Cheng-Dashen point~\cite{cdpoint}, and the 
changes in the $\pi N$ sigma term 
as one goes from the Weinberg point, where the four-momenta of the 
pions is zero, to the Cheng-Dashen point, where the pions are 
on-mass-shell. This question is presently being examined and could shed 
some light on the inconsistencies between the ``observed'' $\Sigma_{N}$ 
and $\sigma_{N}$ as extracted from QCD models.

It is clear from the work presented here that constraints on the coupling
constants and form factors need to be improved before it can be
established whether or not the present potential includes all the
physics of $\pi N$ scattering at low energies.
We are of the opinion that such constraints,
particularly on the coupling constants, should come from QCD or QCD-based 
models.

\acknowledgments

This work was conducted while A. D. L. held an Australian
Postgraduate Award. We acknowledge the financial support
of the Australian Research Council.
We are also indebted to the South Australian Centre for
Parallel Computing for access to their computing facilities.


\newpage


\newpage

\begin{table}
\begin{tabular}{lr} 
cutoff mass & minimum value \\ \hline
$\Lambda _N$        & $m_N + 3 m_{\pi} $     \\
$\Lambda _{\Delta}$ & $m_N + 3 m_{\pi} $     \\
$\Lambda _{\pi}$    & $(m_N + 3 m_{\pi})/2 $ \\
$\Lambda _{\rho}$   & $(m_N + m_{\pi})/2 $   \\
$\Lambda _{\sigma}$ & $(m_N + m_{\pi})/2 $   \\
\end{tabular} 
\vspace*{0.4cm}

\caption{The minimum allowed values of the cutoff masses in terms
of the nucleon and pion masses.}

\label{minall}
\end{table}

\vspace*{1.0cm}

\begin{table}
\begin{tabular}{lrr} 
\mbox{coupling constants} & RS & Pas \\ \hline
$g_{\pi N N}^2/4 \pi$  & 13.5  & 13.5 \\ 
$g_{\pi N N}^{(0)2}/4 \pi$  & 1.80 & 12.1 \\ 
$f_{\pi N \Delta}^2/4 \pi$  &  {\bf 0.365}  & {\bf 0.741} \\
$f_{\pi N \Delta}^{(0)2}/4  \pi$  & {\bf 0.37} & {\bf 0.193} \\ 
$x_{\Delta}$  &    {\bf -0.11}  & --- \\
$g_{\rho}^2/4 \pi$   & {\bf 2.88}  & {\bf 2.73} \\
$\kappa _{\rho}$   & {\bf 2.66}  & {\bf 4.11} \\ 
$g_{\sigma \pi \pi}g_{\sigma NN}/4 \pi$  & {\bf -0.41} & {\bf -3.80} \\
\hline 
\mbox{masses} \\ \hline
$m_N$  &   0.939 & 0.939 \\
$m_N^{(0)}$  & 1.34 & 1.72 \\
$m_{\Delta}$  &  1.232 & 1.232 \\
$m_{\Delta}^{(0)}$   & {\bf 2.305} & {\bf 2.60} \\
$m_{\pi}$ &  0.138 & 0.138 \\
$m_{\rho}$   &  0.769  & 0.769 \\
$m _{\sigma}$  & {\bf 0.65}  & {\bf 0.69} \\ \hline
$\Lambda _N$  & {\bf 3.17} & {\bf 4.90} \\
$\Lambda _{\Delta}$  & {\bf 4.56} & {\bf 3.20} \\
$\Lambda _{\pi}$   & {\bf 1.77} & {\bf 1.76}\\
$\Lambda _{\rho}$   & {\bf 3.67} & {\bf 3.06} \\
$\Lambda _{\sigma}$ & {\bf 1.30} & {\bf 4.26} \\
\end{tabular} 
\vspace*{0.4cm}

\caption{The coupling constants and particle masses obtained using the
Rarita-Schwinger (RS) and Pascalutsa (Pas) $\Delta$ propagators.
The quantities in boldface were varied in the fits.
All masses are in GeV.}
\label{parms}
\end{table}

\begin{table}
\begin{tabular}{lrrrr} 
$\ell _{2I \, 2j}$   & BS (RS)  & BS (Pas)  & \hspace*{1cm} SM95 
& \hspace*{1cm} KH80 \\ \hline
$S_{11}$ &  0.177  &  0.172  &  0.175 &  0.173 \\
$S_{31}$ & -0.101  & -0.105  & -0.087 & -0.101 \\
$P_{11}$ & -0.083  & -0.058  & -0.068 & -0.081 \\
$P_{13}$ & -0.032  & -0.031  & -0.022 & -0.030 \\
$P_{31}$ & -0.041  & -0.041  & -0.039 & -0.045 \\
$P_{33}$ &  0.178  &  0.187  &  0.209 &  0.214 \\ 
\end{tabular} 
\vspace*{0.4cm}

\caption{Scattering lengths and volumes obtained from the
BS equation in units of $m_{\pi} ^{-(2 \ell+1)}$,
compared to results from the SM95~[33] and 
KH80~[58] $\pi N$ partial wave analyses.} 
\label{scatt1}
\end{table}

\vspace*{2.0cm}

\begin{table}
\begin{tabular}{llllll}
$f^2_{\pi N \Delta} / 4 \pi$ & $x_{\Delta}$ & $g^2_{\rho}/4 \pi$ & 
$\kappa _{\rho}$ & equation & ref. \\ \hline
0.365 & -0.11 & 2.88  & 2.66 & BS (RS)      & this work \\
0.741 & --- & 2.73 & 4.26 & BS (Pas) & this work \\
0.35 & -0.3   & 2.5  & 3.7  & ET       & \cite{tjon98} \\
0.43 & ---  & 2.85 & 1.8  & ET       & \cite{tjon98a} \\
0.36 & -0.12 & 3.13 & 2.25 & Sm       & \cite{pj91} \\
0.36 & -0.41 & 2.90 & 1.44 & BbS      & \cite{pj91} \\
0.40 & -0.21 & 3.36 & 6.6  & Ka       & \cite{leeyang94} \\
0.36 & -0.31 & 3.03 & 3.16 & Tr       & \cite{pj91} \\
0.36 & --- & 3.1  & 2.7  & Tr       & \cite{tjon98a} \\ 
\end{tabular}

\vspace*{0.4cm}

\caption{Comparison between coupling constants obtained from the
Bethe-Salpeter (BS), equal-time (ET), smooth (Sm), Blankenbecler-Sugar (BbS),
Kadyshevsky (Ka) equations, and tree-level (Tr) calculations.}
\label{comps}
\end{table}

\begin{table}
\begin{tabular}{lrrrr} 
\mbox{coupling constants} & $n_{\pi}=2$ & $n_{\pi}=4 $
& $n_{\pi}=10$  & model I (RS) \\ \hline
$g_{\pi N N}^2/4 \pi$  & 13.5 & 13.5 & 13.5 & 13.5  \\ 
$g_{\pi N N}^{(0)2}/4 \pi$  & 4.23 & 4.68 & 5.98 & 1.80 \\ 
$f_{\pi N \Delta}^2/4 \pi$  &  {\bf 0.365} &  {\bf 0.365} & {\bf 0.371} & {\bf 0.365} \\
$f_{\pi N \Delta}^{(0)2}/4  \pi$  & {\bf 0.17} & {\bf 0.20} & {\bf 0.196} & {\bf 0.37} \\ 
$x_{\Delta}$  & {\bf -0.13} & {\bf -0.24} & {\bf -0.18}  & {\bf -0.11} \\
$g_{\rho}^2/4 \pi$   & {\bf 2.67}  & {\bf 2.63} & {\bf 2.80} & {\bf 2.88}  \\
$\kappa _{\rho}$   & {\bf 2.18} & {\bf 2.03} & {\bf 2.15} & {\bf 2.66} \\ 
$g_{\sigma \pi \pi}g_{\sigma NN}/4 \pi$  & {\bf 0.86} & {\bf 0.39} & {\bf 0.48} & {\bf -0.41} \\
\hline 
\mbox{masses} \\ \hline
$m_N^{(0)}$  & 1.18 & 1.14 & 1.11 & 1.34 \\
$m_{\Delta}^{(0)}$   & {\bf 1.495} & {\bf 1.492} & {\bf 1.435} & {\bf 2.305} \\
$m _{\sigma}$  & {\bf 0.88} & {\bf 0.62} & {\bf 0.64} & {\bf 0.65} \\ 
$\Lambda _{\pi}$  & {\bf 1.34} & {\bf 1.85} & {\bf 2.73} & {\bf 1.77} \\
\end{tabular} 
\vspace*{0.4cm}

\caption{The coupling constants and particle masses resulting from
fits to the $\pi N$ data using different values of $n_{\pi}$. The parameters
from the model I fit using the Rarita-Schwinger propagator are also shown
for comparison.
The quantities in boldface were varied in the fits. All masses are in GeV. Particle 
masses not given are the same as those given in Table~\ref{parms}.}
\label{pionpars}
\end{table}

\vspace*{2.0cm}

\begin{table}
\begin{tabular}{lrrr} 
\mbox{ } & $n_{\pi }=2$ & $n_{\pi}=4$ & $n_{\pi}=10$ \\ \hline
$\Lambda _{\pi}^R$ & 0.874 & 0.868 & 0.822  \\
$\Delta _{\pi}$ & 2.47\% & 2.51\% & 2.79\%  \\
\end{tabular} 
\vspace*{0.4cm}

\caption{The renormalized pion cutoff masses $\Lambda _{\pi}^R$ (in GeV)
and values of $\Delta _{\pi}$ (expressed as percentages).}
\label{pionren}
\end{table}


\newpage

\newpage

\newpage

\newpage

\end{document}